\documentclass[preprint,showkeys,superscriptaddress,nofootinbib]{revtex4}
\usepackage{amsmath}
\usepackage{amssymb}
\usepackage{booktabs}

\usepackage{here}

\usepackage[pdftex]{graphicx,color}
\usepackage{epstopdf}
\usepackage[bookmarks=true,bookmarksnumbered=true,bookmarkstype=toc]{hyperref} 

\allowdisplaybreaks[4]
\newcommand{\lsim}{\raise0.3ex\hbox{$\;<$\kern-0.75em\raise-1.1ex\hbox{$\sim\;$}}}
\newcommand{\gsim}{\raise0.3ex\hbox{$\;>$\kern-0.75em\raise-1.1ex\hbox{$\sim\;$}}}

\newcommand{\sign}[1]{\mathrm{sign}\big[#1\big]}

\preprint{KUNS-2389}
\preprint{YITP-12-17}

\begin{document}

%%%%% declaration for front matter%%%%%%%%%%%%%%%%%%%%%%%%%%%%%%%%%%%
\title{Constraining the Higgs sector from False Vacua 
in the Next-to-Minimal Supersymmetric Standard Model}

\author{Tatsuo Kobayashi}
\email{kobayash@gauge.scphys.kyoto-u.ac.jp}
\affiliation{Department of Physics, Kyoto University, Kyoto 606-8502, Japan}

\author{Takashi Shimomura}
\email{stakashi@muse.sc.niigata-u.ac.jp}
\affiliation{Yukawa Institute for Theoretical Physics, Kyoto University, Kyoto 606-8502, Japan}
\affiliation{Department of Physics, Niigata University,~Niigata, 950-2181, Japan}

\author{Tsubasa Takahashi}
\email{tsubasa@yukawa.kyoto-u.ac.jp}
\affiliation{Yukawa Institute for Theoretical Physics, Kyoto University, Kyoto 606-8502, Japan}

\date{\today}

%\preprint{YITP-11-25}
%\preprint{KUNS-2323}

%%%%% typeset front matter (including abstract) %%%%%%%%%%%%%%%%%%%%%%
\begin{abstract}
We study the mass, the mixing and the coupling with $Z$ boson of the lightest Higgs boson 
in the next-to-minimal supersymmetric standard model. The vacuum structure of the Higgs potential is 
analyzed and the new false vacua are discussed.
The significant parameter region can be excluded by requiring that the realistic vacuum is deeper than false vacua, 
which result in constraints on the properties of the lightest Higgs boson.
\end{abstract}

\maketitle
%%%%%%%%%%%%%%%%%%%%%%
\section{Introduction}\label{sec:introduction}
%%%%%%%%%%%%%%%%%%%%%%
Supersymmetric extension of the standard model is one of promising candidates 
for physics beyond the weak scale.
In particular, the minimal supersymmetric standard model (MSSM) is 
interesting.
However, the MSSM has the so-called $\mu$ problem \cite{Kim:1983dt}.
The $\mu$-term is the supersymmetric mass term 
of Higgs fields.
In addition, Higgs fields have soft scalar mass term due to 
supersymmetry (SUSY)   breaking.
To realize the successful electroweak symmetry breaking (EWSB), 
both sizes of the $\mu$ and soft scalar mass must be 
of the same order.
Why can the two masses with different sources be 
of the same order ?
That is the $\mu$ problem.
Furthermore, the LHC bounds $m_h > 115$ GeV \cite{:2012si,Chatrchyan:2012tx}
also give a significant constraint on 
the Higgs sector of the MSSM.

The next-to minimamal supersymmetric standard model (NMSSM) 
is the simplest extenstion of the MSSM by adding a singlet field $S$ 
\cite{Fayet:1974pd,Fayet:1976et,Fayet:1977yc,Fayet:1979sa,Nilles:1982dy,
Frere:1983ag,Derendinger:1983bz,Ellis:1988er,Drees:1988fc}
( for a review, see \cite{Ellwanger:2009dp}) .
The $\mu$-term is forbidden by the $\mathbb{Z}_3$ discrete symmetry, 
but it is effectively induced though the coupling 
$\lambda SH_1 H_2$ in the superpotential after the scalar component of 
$S$ developes its vacuum expectation value (vev).
Furthermore, such a vev is determined by SUSY breaking 
terms.
Thus, the vev is related with the size of SUSY breaking, 
and the $\mu$-problem can be resolved.

In addition, the Higgs sector of the NMSSM has 
significantly different aspects from one in the MSSM 
(see e.g. \cite{Ellwanger:2009dp,Ellwanger:2011sk}).
The lightest Higgs mass at the tree-level can be 
larger than one in the MSSM, because the above 
coupling term $\lambda SH_1 H_2$ in the superpotential 
leads to a new quartic term of $H_1$ and $H_2$ in 
the Higgs potential.
The larger value of $\lambda$ would increase 
the lightest Higgs mass.
Moreover, the Higgs fields $H_1$ and $H_2$ are 
mixed with the singlet $S$ after the symmetry breaking.
Such mixing changes the coupling between the Higgs scalars and 
 vector bosons.
Thus, the Higgs sector of the NMSSM has a rich 
structure.

Furthermore, the behavior of the Higgs potential  
and its vacuum structure are much more complicated 
in the NMSSM than those in the MSSM.
The Higgs potential of the NMSSM should include 
the realistic minimum, where the successful EWSB is realized.
In addition to the realistic vacuum, 
the Higgs potential may include other (local) minima, some of which 
do not break the EW symmetry correctly.
If such false vacua are deeper than the realistic vacuum, 
the realistic vacuum may not be realized.\footnote{
In addition, there may be wrong vacua, where 
squarks and sleptons develop their vevs.
On such vacua, charge and/or color are broken \cite{Frere:1983ag,AlvarezGaume:1983gj,Derendinger:1983bz,
Kounnas:1983td,Claudson:1983et,Drees:1985ie,Gunion:1987qv,Komatsu:1988mt,Gamberini:1989jw,Casas:1995pd,Kobayashi:2010zx,Kanehata:2010ci}.}
Hence, our parameter space of SUSY breaking terms and 
dimensionless couplings in the NMSSM would be constrained 
in order to avoid such false vacua 
such that false vacua are less deeper than the realistic vacuum.
In fact, numerical studies have been done 
and also analytical studies along certain directions 
have been carried out \cite{Ellwanger:1996gw,Funakubo:2004ka,Cerdeno:2004xw,Maniatis:2006jd,Cerdeno:2007sn,Franceschini:2010qz,Cerdeno:2011qv,Kanehata:2011ei,Bertuzzo:2011ij,Hamaguchi:2011kt}.
Then, it was shown that significant parameter regions can be excluded 
by requirement to avoid false vacua.
For example, the NMSSMTools\footnote{see http://www.th.u-psud.fr/NMHDECAY/nmssmtools.html} is the famous code 
to analyze several phenomenological aspects 
in the NMSSM \cite{Ellwanger:2005dv,Ellwanger:2006rn,Djouadi:2008uw}, 
and it includes some of analytical conditions 
to avoid certain false vacua.

Our purpose in this paper is to study phenomenological 
aspets of the Higgs sector in the NMSSM such as 
the physical mass spectrum of Higgs scalars and their mixing 
with taking into account analytical conditions to 
avoid false vacua.
As false vacua, we consider directions, which are new 
and not included, e.g. in the NMSSMTools.
Then, we will show that important parameter regions 
could be excluded by such conditions.
Obviously, the parameter region, where 
the Higgs masses are tachyonic on the realistic vacuum, 
is excluded.
When the doublet Higgs and singlet Higgs scalars mix 
sizablely, the lightest Higgs boson mass may become tachyonic.
When there are such tachyonic modes, there would be 
a wrong vacuum deeper than the realistic vacuum.
Thus, the parameter region with the tachyonic Higgs boson mass 
corresponds to the region, where false vacua are 
deeper than the realistic vacuum.
Furthermore, the parameter region leading to 
deeper false vacua would be near and outside  
the region with the tachyonic Higgs boson mass on 
the realistic vacuum.
Thus, wider regions of the parameter space, 
in particular the parameter regions with sizable mixing between 
doublet and singlet Higgs scalars would be 
excluded by requiring to avoid false vacuum.

This paper is organized as follows.
In section \ref{sec:real-vacu-higgs}, we review 
the Higgs sector of the NMSSM, in particular 
the realistic vacuum and Higgs boson masses.
In section \ref{sec:false-vacua-along}, 
we study new false vacua, which can be deeper than the 
realistic vacuum.
In section \ref{sec:numerical-analysis}, we study numerically 
implications of our constraints.
Section \ref{sec:conclusion} is devoted to conclusions and 
discussions.

%%%%%%%%%%%%%%%%%%%%%%%%%%%%%%%%%%%%%%%%%%%%%%%%%%%%%%%%%%%%%
\section{realistic vacuum and the Higgs masses in the NMSSM.}\label{sec:real-vacu-higgs}
%%%%%%%%%%%%%%%%%%%%%%%%%%%%%%%%%%%%%%%%%%%%%%%%%%%%%%%%%%%%%
We start our discussion with briefly reviewing the realistic vacuum and the masses of the Higgs bosons in the NMSSM.
The NMSSM is defined by adding a gauge singlet chiral supermultiplet $\hat{S}$ and imposing a global 
$\mathbb{Z}_3$ symmetry to the MSSM. Due to the $\mathbb{Z}_3$ symmetry, the superpotential consists of only terms involving 
three chiral supermultiplets, thus dimensionful couplings as a supersymmetric Higgsino mass term and the tad pole term 
are forbidden. In the following, fields with a hat $(\hat{~})$ symbol represent superfields and those without the symbol 
represent the corresponding scalar fields. The superpotential of the Higgs and the singlet superfields is given by 
\begin{align} 
  \mathcal{W}_{\mathrm{Higgs}} &= - \lambda\hat{S} \hat{H}_1\cdot\hat{H}_2 + \frac{1}{3}\kappa\hat{S}^3, \label{eq:1}
\end{align}
where $\lambda$ and $\kappa$ are the Yukawa coupling constants of the Higgs fields, and $\hat{H}_1$ and $\hat{H}_2$ are 
the down-type and the up-type Higgs supermultiplets defined as
\begin{align}
 \hat{H}_1 =
  \begin{pmatrix}
   \hat{H}_1^0 \\
   \hat{H}_1^-
  \end{pmatrix},~~
\hat{H}_2 =
  \begin{pmatrix}
   \hat{H}_2^+ \\
   \hat{H}_2^0
  \end{pmatrix},\label{eq:38} 
\end{align} 
respectively.

The realistic vacuum which breaks the electroweak (EW) symmetry successfully can be found by minimizing the Higgs potential.
The Higgs potential is obtained from $F$-, $D$-terms and 
the soft SUSY breaking terms given by
\begin{align}
  V_{\mathrm{soft}} & = m^2_{H_1} H_1^\dagger H_1 + m^2_{H_2} H_2^\dagger H_2 + m^2_{S} S^\dagger S 
                      - \left( \lambda A_\lambda S H_1 \cdot H_2 
                       - \frac{1}{3} \kappa A_\kappa S^3 + h.c.\right), \label{eq:2}
\end{align}
where $m^2_{H_1,H_2,S}$ and $A_{\lambda,\kappa}$ are soft masses and trilinear couplings of the scalars, 
respectively. For the EW symmetry to be successfully broken, the neutral Higgs fields 
develop vevs while vevs of the charged Higgs fields are vanishing. 
Using the gauge transformations, without loss of generality, one can take $\langle H_2^+ \rangle = 0$ and 
$\langle H_2^0 \rangle \in \mathbb{R}^+$. The condition for vanishing $\langle H_1^- \rangle$ is to require that 
the charged Higgs scalars have positive mass squareds. Then, the potential of the neutral Higgs fields is given by,
\begin{align}
 V =& \lambda^2|S|^2 \left(|H_1^0|^2 + |H_2^0|^2 \right) + |F_S|^2 + V_D \nonumber \\
    & + m_{H_1}^2 |H_1^0|^2 + m_{H_2}^2 |H_2^0|^2 + m_{S}^2 |S|^2 
    -  (\lambda A_\lambda H_1^0 H_2^0 S - \frac{1}{3} \kappa A_\kappa S^3 + h.c.), \label{eq:3}
\end{align}
where $F_S$ and $V_D$ denote the $F$-term of $\hat{S}$ and $D$-term potential,
\begin{align}
 F_S^\ast &= \kappa S^2 - \lambda H^0_1 H^0_2, \label{eq:8} \\
 V_D &= \frac{1}{8} (g_1^2 + g_2^2) \left(|H_1^0|^2 - |H_2^0|^2 \right)^2. \label{eq:9}
\end{align}
Here, $g_1$ and $g_2$ denote the gauge coupling constants of $U(1)_Y$ and $SU(2)_L$, respectively.
The Higgs sector of the NMSSM is characterized by the following parameters,
\begin{align}
\lambda, ~~\kappa, ~~m_{H_1}^2, ~~m_{H_2}^2, ~~m_{S}^2, ~~A_\lambda ~~{\rm and} ~~A_\kappa.
\label{eq:parameter-1}
\end{align}

In the following discussions, we assume that all of soft masses, trilinear couplings and Yukawa couplings 
are real for simplicity. Although the vevs of $H_1^0$ and $S$ can be complex in general under this assumption,
it was shown in \cite{Romao:1986jy} that such CP violating extrema are maxima rather than minima. Thus, it is reasonable to assume that 
the neutral Higgs fields develop real and non-vanishing vevs while the charged ones do not. Then, we denote vevs as 
\begin{align}
\langle H_1^0 \rangle = v_1, \qquad \langle H_2^0 \rangle = v_2, 
\qquad \langle S \rangle = s. \label{eq:39}
\end{align}
Furthermore, as was discussed in \cite{Cerdeno:2004xw}, the Higgs potential, (\ref{eq:3}), is invariant under the replacements, 
$\lambda,~\kappa,~s \rightarrow -\lambda,~-\kappa,~-s$ and $\lambda,~v_1 \rightarrow -\lambda,~-v_1$, 
therefore we can always take $\lambda$ and $v_1$ to be positive while $\kappa,~\mu (\equiv \lambda s)$ and 
$A_\lambda,~A_\kappa$ can have both signs. 
The exsitence of the minima of the Higgs potential is classified according to the signs of $\kappa$, $s$ and 
$A_\lambda$ and $A_\kappa$ 
\cite{Cerdeno:2004xw},
\begin{enumerate}
 \item for positive $\kappa$,
       \begin{enumerate}
	\item when $\sign{s} = \sign{A_\lambda} = -\sign{A_\kappa}$, the minima always exist.
	\item when $\sign{s} = -\sign{A_\lambda} = -\sign{A_\kappa}$, \\
	      the minima exist if $|A_\kappa| > 3\lambda v_1 v_2 |A_\lambda|/(- |s A_\lambda| + \kappa |s^2|)$ 
	      where the denominator is positive.
	\item when $\sign{s} = \sign{A_\lambda} = \sign{A_\kappa}$, \\
	      the minima exist if $|A_\kappa| < 3\lambda v_1 v_2 |A_\lambda|/(|s A_\lambda| + \kappa |s^2|)$.
       \end{enumerate}
 \item for negative $\kappa$
       \begin{enumerate}
	\item[(d)] when $\sign{s} = \sign{A_\lambda} = \sign{A_\kappa}$, \\
		   the minima exist if $|A_\kappa| < 3\lambda v_1 v_2 |A_\lambda|/(|s A_\lambda| - \kappa |s^2|)$ 
		   where the denominator is positive.
       \end{enumerate}
\end{enumerate}
The vevs are determined by the stationary conditions or minimizing the potential, (\ref{eq:3}), 
with respect to the neutral Higgs fields,
\begin{subequations}
 \begin{align}
   \frac{\partial V}{\partial H^0_1} &= \lambda^2 v \cos\beta ( s^2 + v^2 \sin^2 \beta )
     - \lambda \kappa v s^2 \sin\beta + \frac{1}{4} g^2 v^3 \cos\beta \cos2 \beta \nonumber \\
    &\quad + m_{H_1}^2 v \cos\beta - \lambda A_\lambda v s \sin\beta = 0, \label{eq:20} \\
   \frac{\partial V}{\partial H^0_2} &= \lambda^2 v \sin\beta ( s^2 + v^2 \cos^2 \beta )
     - \lambda \kappa v s^2 \cos\beta - \frac{1}{4} g^2 v^3 \sin\beta \cos2 \beta \nonumber \\
    &\quad + m_{H_2}^2 v \sin\beta - \lambda A_\lambda v s \cos\beta = 0, \label{eq:21} \\
   \frac{\partial V}{\partial S} &= \lambda^2 s v^2 + 2 \kappa^2 s^3 - \lambda \kappa v^2 s \sin 2\beta
    + m_S^2 s - \frac{1}{2} \lambda A_\lambda v^2 \sin 2\beta + \kappa A_\kappa s^2 = 0, \label{eq:22}
 \end{align}
\label{eq:4}
\end{subequations}
where $v = \sqrt{v_1^2 + v_2^2}$, $\tan\beta = v_2/v_1$ and $g^2 = g_1^2 + g_2^2$. The vevs of the doublet Higgs fields must satisfy 
$v \simeq 174$ GeV to give the correct masses to the gauge bosons. 
Without special relations among parameters, when two of the Higgs fields are non-vanishing, 
the other must be non-vanishing, too. Therefore, a non-trivial solution of (\ref{eq:4}) is as follows; either three Higgs fields 
are non-vanishing or one Higgs field is non-vanishing \cite{Kanehata:2011ei}.
This fact is originated from the trilinear terms, $\lambda A_\lambda H_1^0 H_2^0 S$, 
in the soft SUSY breaking terms and the quartic term, $\lambda \kappa H_1^0 H_2^0 (S^\ast)^2$, 
in the $F$-term potential.
This observation justifies our strategy of analyses on false minima of the Higgs potential in the next section.

It is useful to express the soft SUSY breaking masses in terms of other parameters rewriting the stationary conditions, (\ref{eq:4}), 
\begin{subequations}
\begin{align}
m_{H_1}^2 &= - \mu^2 - \frac{2 \lambda^2}{g^2} m_Z^2 \sin^2\beta - \frac{1}{2} m_Z^2 \cos 2\beta 
           + \mu \left( \frac{\kappa}{\lambda} \mu + A_\lambda \right) \tan \beta, \label{eq:11}\\
m_{H_2}^2 &= - \mu^2 - \frac{2 \lambda^2}{g^2} m_Z^2 \cos^2\beta + \frac{1}{2} m_Z^2 \cos 2\beta 
           + \mu \left( \frac{\kappa}{\lambda} \mu + A_\lambda \right) \cot \beta, \label{eq:12}\\
    m_S^2 &= - \frac{2 \lambda^2}{g^2}m_Z^2 - \frac{2 \kappa^2}{\lambda^2} \mu^2 + \frac{2 \lambda \kappa}{g^2} m_Z^2 \sin 2\beta 
           + \frac{\lambda^2}{g^2} \frac{A_\lambda m_Z^2}{\mu} \sin 2\beta - \frac{\kappa}{\lambda} A_\kappa \mu, \label{eq:13}
\end{align}
\label{eq:7}
\end{subequations}
where $m_Z^2 = \frac{1}{2} g^2 v^2$ and $\mu = \lambda s$.
Thus, given $m_Z$, we can use the following parameters,
\begin{align}
\lambda, ~~\kappa, ~~A_\lambda, ~~A_\kappa,~~\tan \beta 
~~{\rm and} ~~\mu,
\label{eq:parameter-2}
\end{align}
instead of (\ref{eq:parameter-1}).
Using these parameters, the minimum of the realistic vacuum, which reproduces the observed $Z$ boson mass, can be written as
\begin{align}
 V_\mathrm{min} &=
- \lambda^2 \frac{m_Z^4 \sin^2 2 \beta}{g^4} - \frac{m_Z^4 \cos^2 2 \beta}{2 g^2} 
+ \overline{V}^{S}_{\mathrm{min}}, \label{eq:6}
\end{align}
where $\overline{V}^S_{\mathrm{min}}$ is the potential involving only $s=\mu/\lambda$, 
\begin{align}
 \overline{V}^S_{\mathrm{min}} = 
\frac{\kappa^2}{\lambda^4}  \mu^4
+ \frac{2}{3} \frac{\kappa}{\lambda^3} A_\kappa \mu^3
+ \frac{1}{\lambda^2} m_S^2 \mu^2, \label{eq:5}
\end{align}
with $m_S^2$ given by \eqref{eq:13}.
In the following section, we study false vacua and compare their depths with (\ref{eq:6}).

The mass-squared matrices of the Higgs bosons at tree-level are obtained from (\ref{eq:3}) by expanding the Higgs 
fields around their vevs. The number of degrees of freedom of the Higgs bosons is ten, and three of them are absorbed by gauge bosons via the Higgs mechanism. The remaining seven physical degrees correspond to three CP-even Higgs bosons, two CP-odd Higgs bosons 
and one charged Higgs boson. The mass-squared matrix of the CP-even Higgs bosons is real-symmetric and denoted as $M_{h,ij}^2$ 
where $i,~j$ runs over $1$ to $3$ for the down-type, the up-type and the singlet Higgs scalars. It is given by 
\begin{subequations}\label{eq:23}
 \begin{align}
  M_{h,11}^2 &= m_Z^2 \cos^2 \beta + \mu \left(\frac{\kappa}{\lambda} \mu + A_\lambda \right) \tan \beta, \label{eq:15}\\
  M_{h,22}^2 &= m_Z^2 \sin^2 \beta + \mu \left(\frac{\kappa}{\lambda} \mu + A_\lambda \right) \cot \beta, \label{eq:16}\\
  M_{h,33}^2 &= \frac{4 \kappa^2}{\lambda^2} \mu^2 + \frac{\kappa}{\lambda} A_\kappa \mu 
              + \frac{\lambda^2}{g^2} \frac{A_\lambda m_Z^2}{\mu} \sin 2\beta, \label{eq:17} \\
  M_{h,12}^2 &= 2 \left( \frac{\lambda^2}{g^2} - \frac{1}{4} \right) m_Z^2 \sin 2\beta 
              - \mu \left(\frac{\kappa}{\lambda} \mu + A_\lambda \right), \label{eq:40} \\
  M_{h,13}^2 &= \frac{2\sqrt{2} \lambda}{g} \mu m_Z \cos\beta 
              - \frac{\sqrt{2} \lambda}{g} m_Z \left( A_\lambda + \frac{2 \kappa}{\lambda} \mu \right) \sin\beta, \label{eq:41} \\
  M_{h,23}^2 &= \frac{2\sqrt{2} \lambda}{g} \mu m_Z \sin\beta 
              - \frac{\sqrt{2} \lambda}{g} m_Z \left( A_\lambda + \frac{2 \kappa}{\lambda} \mu \right) \cos\beta. \label{eq:42}
 \end{align}
\end{subequations}
The mass-squared matrix of the CP-odd Higgs bosons, $M_A^2$, is also real-symmetric and given by
\begin{subequations}\label{eq:24}
 \begin{align}
  M^2_{A,11} &= \frac{2 \mu}{\sin 2\beta} \left( A_\lambda + \frac{\kappa}{\lambda} \mu \right), \label{eq:18}\\
  M^2_{A,22} &= \frac{\lambda^2}{g^2} m_Z^2 \left( \frac{A_\lambda}{\mu} + \frac{4 \kappa}{\lambda} \right) \sin 2\beta 
              - \frac{3 \kappa}{\lambda} A_\kappa \mu, \label{eq:19}\\
  M^2_{A,12} &= \frac{\sqrt{2} \lambda}{g} m_Z \left( A_\lambda - \frac{2 \kappa}{\lambda} \mu \right), \label{eq:43}
 \end{align}
\end{subequations}
where we have removed the Nambu-Goldstone mode.

It can be shown that the mass of the lightest CP-odd Higgs boson vanishes when $\kappa$ goes to zero. This is because the Peccei-Quinn 
symmetry is restored in this limit. Hence $\kappa$ should not be much small to avoid tachyonic CP-odd Higgs bosons.
Assuming $|\mu| > m_Z$ to be consistent with non-observation of the charged Higgsinos, 
we can derive intuitive conditions to avoid tachyonic Higgs bosons from (\ref{eq:23}) and (\ref{eq:24}).
Firstly from (\ref{eq:15}), (\ref{eq:16}) and (\ref{eq:17}), it can be understood that the mass 
of the lightest CP-even Higgs tends to become tachyonic if the 
following conditions are satisfied,
\begin{align}
 \mu \left(\frac{\kappa}{\lambda} \mu + A_\lambda \right) \ll -m_Z^2, \label{eq:26}
\end{align}
and/or
\begin{align}
 \frac{\kappa}{\lambda} A_\kappa \mu \ll -m_Z^2. \label{eq:27}
\end{align}
Necessary conditions to avoid the tachyonic Higgs bosons are to require the left-hand side of \eqref{eq:26} and \eqref{eq:27} 
to be positive, i.e.
\begin{enumerate}
 \item For positive $\kappa$, $A_\lambda \mu$ and $A_\kappa \mu$ should be positive.
 \item For negative $\kappa$, $A_\kappa \mu$ should be negative and $\lambda A_\lambda \mu > -\kappa \mu^2$ should be satisfied.
\end{enumerate}
On the other hand, when $\kappa$ is positive (negative), we can expect from (\ref{eq:19}) that $A_\kappa \mu$ should be 
negative (positive) to avoid for the lightest CP-odd Higgs boson to be tachyonic. 
Thus there is a tension between the conditions for non-tachyonic 
modes of the CP-even and the CP-odd Higgs bosons. This tension can be avoided if magnitudes of $A_\lambda$ and $A_\kappa$ 
should be tuned so that both conditions are satisifed. One of the choices of $A_\lambda$ and $A_\kappa$ for positive $\kappa$
is obtained as \cite{Ellwanger:2006rm}
\begin{align}
A_\lambda \mu > -\frac{k}{\lambda}\mu^2 ~\mathrm{and}~-\frac{4 \kappa^2}{\lambda^2} \mu^2 < \frac{\kappa}{\lambda} A_\kappa \mu < 0,\label{eq:30}
\end{align}
where tachyonic Higgs bosons can be also avoided when $\kappa$ is
negative.
% \red{(Is this valid if $\kappa$ is negatively large?)}. 
The condition \eqref{eq:30} is enough condition stating that parameters not satisfying this condition result in the tachyonic masses 
of the Higgs bosons. This condition is useful to understand the behavior of the tachyonic mass region. 
As we will show in \ref{sec:numerical-analysis}, the tachyonic mass
region appears near the regions given by \eqref{eq:30}. 
The conditions imply that 
the effective Higgsino mass, $|\mu|$, should be larger than $|A_\kappa|$ .
%\red{(Is this valid if $\lambda$ is very small and/or $\kappa$ is very large?)}. 
Since $\mu$ is of order TeV scale not to introduce the little 
hierarchy problem, $|A_\kappa|$ must be relatively small.

Furthermore, we can derive other conditions by taking into account 
the off-diagonal terms. The mass of the lightest CP-even Higgs boson tends to become tachyonic when the mixings between the doublet Higgs 
and the singlet bosons are large. The mixings can be naively written as
\begin{align}
 2 \lambda \mu - (\lambda A_\lambda + 2 \kappa \mu) \sin2\beta.\label{eq:28}
\end{align}
Requiring this mixing to be vanishing, we have the condition \cite{Ellwanger:2006rm}
\begin{align}
 A_\lambda \simeq \frac{2\mu}{\sin 2\beta} - 2 \frac{\kappa}{\lambda} \mu.\label{eq:29}
\end{align}
Thus the tachyonic Higgs boson tends to appear when the left and right hand sides are not comparable. 
As we mentioned, the supersymmetric Higgsino mass should be about TeV scale. This implies that $A_\lambda$ must be 
taken to be relatively large due to the factor $2/\sin 2\beta$. One might consider that small $A_\lambda$ 
can be taken when $\kappa/\lambda$ is 
larger than one so that the left-hand side of \eqref{eq:29} is tuned. In such parameter region, 
however, tachyonic Higgs bosons do not appear even if 
$A_\lambda$ is larger than the right-handed side of \eqref{eq:29}.
Indeed, as we will see in Sec.~\ref{sec:numerical-analysis}, the tachyonic Higgs boson appears for 
small values of $|\kappa/\lambda|$. This is because the element of the mass-squared, (\ref{eq:17}), 
increases and becomes larger than the off-diagonal element (\ref{eq:41}) and (\ref{eq:42}) 
as $|\kappa/\lambda|$ increases.
The mixing of the singlet in the lightest Higgs boson is suppressed in this region.

The mass squared of the charged Higgs boson is 
\begin{align}
 m^2_{H^\pm} = m_W^2 - \frac{2 \lambda^2}{g^2} m_Z^2 + \frac{2 \mu}{\sin 2\beta} \left(A_\lambda + \frac{\kappa}{\lambda} \mu \right), \label{eq:25}
\end{align}
where $m_W^2 = \frac{1}{2}g_2^2 v^2$ is the mass squared of the $W$ boson. 
The charged Higgs boson mass squared can also be tachyonic when $\lambda$ is large enough.
These mass-squared matrices are used in numerical calculations to find tachyonic mass regions.

%%%%%%%%%%%%%%%%%%%%%%%%%%%%%%%%%%%%%%%%%%%%%%%
\section{False vacua along specific directions} \label{sec:false-vacua-along}
%%%%%%%%%%%%%%%%%%%%%%%%%%%%%%%%%%%%%%%%%%%%%%%
In this section, we show that false vacua in the Higgs potential can be found by considering specific directions. 
Hereafter we analyze the Higgs potential involving only the neutral Higgs fields.  
As discussed in the previous section,
when two of the Higgs fields develop their vevs, the other must develop its vev to satisfy the stationary conditions.
Thus, analyses of the Higgs potential are constrained to cases of
either one or three non-vanishing Higgs fields.
Obviously, the vacua, where only one of Higgs fields has
non-vanishing vev, are unrealistic, and the conditions to avoid such 
vacua have been studied in \cite{Ellwanger:1996gw,Kanehata:2011ei}.
The realistic vacuum given in \eqref{eq:4} is included along the direction where all of three Higgs fields develop their vevs.
The cases with three non-vanishing Higgs vevs also include false vacua on which the EWSB does not occur correctly. 
Analyses with three non-vanishing Higgs fields are so complicated in general 
that it can not be performed analytically. However, analytical study is possible for specific directions in field space along which 
some of the Higgs fields are related or vanishing. In general, minima of the scalar potential appear when positive quartic terms balance 
 with negative quadratic
and trilinear terms.  When the quartic terms in
the scalar potential, (\ref{eq:3}), vanish, the false minima appear for large values of the Higgs fields and hence these become deeper.
Thus we restrict our discussions to three possible cases in which 
three Higgs fields are aligned so that $D$-term and/or $F_S$-term are vanishing.
In \cite{Kanehata:2011ei}, false vacua with $|H_1| = |H_2| \neq 0$ and  $S\neq 0$, which corresponds to $F_S=0$ and $V_D=0$, was studied.
In fact,  false vacua deeper than the realistic vacuum are easily found 
along these directions. Such directions should be avoided to stabilize the realistic minimum. 
In the following, we denote the neutral Higgs fields $H_{1,2}^0$ as $H_{1,2}$ for abbreviation.

%%%%%%%%%%%%%%%%%%%%%%%%%%%
\subsection{$F_S = V_D =0$ direction}\label{sec:f_s=v_d=0}
%%%%%%%%%%%%%%%%%%%%%%%%%%%
Fisrt we consider a direction where both of $F_S$ and $V_D$ are vanishing. The vacua along this direction were firstly 
analyzed in \cite{Kanehata:2011ei} and shown that those can be deeper than the realistic vacuum. Along this direction, 
referred as $A$, the Higgs fields 
are not independent and related as 
\begin{subequations}
\begin{align}
 H_1 &= H_2,\label{eq:33}\\
 S^2 &= \frac{\lambda}{\kappa} H_1 H_2.\label{eq:31}
\end{align}
\end{subequations}
Here positive values of $H_1$ and $H_2$ are assumed since we are interested in the false vacua near the realistic vacua. 
Then, $\kappa$ must be positive to satisfy Eq.~(\ref{eq:31}). When we require perturvativity up to the GUT scale 
($\simeq 10^{16}$ GeV), i.e. all of the gauge and Yukawa couplings to be smaller than $\sqrt{4\pi}$ below the GUT scale, 
the minimum appearing along the direction is unacceptable. This is because the Yukawa couplings encounter the Landau pole before 
the GUT scale when $\tan\beta = v_2/v_1 = 1$ \cite{Kanehata:2011ei}. We will show this numerically in the next section.

The scalar potential along this direction can be written as
\begin{align}
 V_A = \hat F {H_2}^4 - 2 \hat A {H_2}^3 + \hat m^2 {H_2}^2, \label{eq:34}
\end{align}
where
\begin{subequations}
\begin{align}
 & \hat{F} = \frac{2 \lambda^3}{\kappa}, \label{eq:44} \\
 & \hat{A} = \lambda \sqrt{\frac{\lambda}{\kappa}} 
              \left( \epsilon_1 |A_\lambda| - \frac{1}{3} \epsilon_2 |A_\kappa| \right), \label{eq:45}\\
 & \hat{m}^2 = m_{H_1}^2 + m_{H_2}^2 +  \frac{\lambda}{\kappa} m_S^2, \label{eq:46}
\end{align}
\label{eq:49}
\end{subequations}
and
\begin{subequations}
\begin{align}
 \epsilon_1 &\equiv \sign{A_\lambda H_1 H_2 S} = \sign{A_\lambda} \sign{S}, \label{eq:47} \\
 \epsilon_2 &\equiv \sign{\kappa A_\kappa S^3} = \sign{A_\kappa} \sign{S}. \label{eq:48}
\end{align}
\label{eq:36}
\end{subequations}
The minimum of the potential becomes 
deeper when the trilinear term, $\hat{A}$, is positive. From Eqs.~(\ref{eq:36}), the trilinear term can be always taken to 
be positive using the sign of $S$ and is given as
\begin{align}
 \hat{A} = \lambda \sqrt{\frac{\lambda}{\kappa}} \left| A_\lambda - \frac{1}{3} A_\kappa \right|.\label{eq:50}
\end{align}
By minimizing the potential of (\ref{eq:34}) with respect to $H_2$, the value of $H_2$ at extremal 
is obtained as
\begin{align}
 {H_2}_{\mathrm{ext}} = \frac{3 \hat{A}}{4 \hat{F}}
                       \left( 1 + \sqrt{1 - \frac{8 \hat{m}^2 \hat{F}}{9 \hat{A}^2} }~ \right),\label{eq:10}
\end{align}
where $\hat{m}^2 \le \frac{9 \hat{A}^2}{8 \hat{F}}$ is required for ${H_2}_{\mathrm{ext}}$ to be real.
Then, the minimum of the potential is obtained by inserting (\ref{eq:10}) into \eqref{eq:34} as
\begin{align}
 V_{A,\mathrm{min}} = -\frac{1}{2} {H_2}_{\mathrm{ext}}^2 (\hat{A} {H_2}_{\mathrm{ext}} - \hat{m}^2).\label{eq:14}
\end{align}
To realize the correct EWSB, the following necessary condition is required,
\begin{align}
 V_{A,\mathrm{min}} \ge V_{\mathrm{min}}. \label{eq:51}
\end{align}

%%%%%%%%%%%%%%%%%%%%%%%%%%%%%%%%%%%%%%
\subsection{$F_S \neq 0$ and $V_D =0$ direction}\label{sec:f_s-neq-0_and_v_d=0}
%%%%%%%%%%%%%%%%%%%%%%%%%%%%%%%%%%%%%%
Next, we analyze the direction where $V_D$ is vanishing while $F_S$ is non-vanishing. We refer this direction as 
the direction $B$. 
The vevs of the up-type and down-type Higgs scalars must always satisfy the relation
\begin{align}
 H_1 = H_2, \label{eq:52}
\end{align}
for $V_D$ to be vanishing. Similar to \ref{sec:f_s=v_d=0}, the Yukawa couplings blow up before the GUT scale, and 
therefore the minima appearing along the direction should be avoided.

We parametrize the vev of $S$ as 
\begin{align}
 S = \sign{S} \beta H_2, \label{eq:53}
\end{align}
where $\beta$ is positive by definition. Then the potential is given by
\begin{align}
 V_B = \hat{F} {H_2}^4 -2 \hat{A} {H_2}^3 + \hat{m}^2 {H_2}^2, \label{eq:54}
\end{align}
where
\begin{subequations}
\begin{align}
 & \hat{F} = \kappa^2 \beta^4 + \lambda^2 ( 1 + 2 \beta^2 ) -2 \epsilon_1 \lambda |\kappa| \beta^2, \label{eq:55} \\
 & \hat{A} = \left( \epsilon_2 \lambda |A_\lambda| - \frac{1}{3} \epsilon_3 |\kappa| |A_\kappa| \beta^2 \right) \beta, \label{eq:56} \\
 & \hat{m}^2 = m_{H_1}^2 + m_{H_2}^2 +  m_S^2 \beta^2. \label{eq:57}
\end{align}
\label{eq:61}
\end{subequations}
and 
\begin{subequations}
\begin{align}
 \epsilon_1 &= \sign{\kappa H_2 (S^\ast)^2} = \sign{\kappa}, \label{eq:58} \\
 \epsilon_2 &= \sign{A_\lambda H_2 S} = \sign{A_\lambda} \sign{S}, \label{eq:59} \\
 \epsilon_3 &= \sign{\kappa A_\kappa S^3} = \sign{\kappa} \sign{A_\kappa} \sign{S}. \label{eq:60}
\end{align}
\label{eq:37}
\end{subequations}
We can expect that the deepest direction will 
be found along the positive trilinear terms with $\kappa > 0$ and $A_\lambda A_\kappa <0$ so 
that $\hat{F}$ becomes smaller and $\hat{A}$ becomes larger.
From Eqs.~(\ref{eq:37}), the trilinear term can be 
always taken to be positive, and the quartic and the trilinear terms are given by
\begin{subequations}
\begin{align}
 & \hat{F} = \kappa^2 \beta^4 + \lambda^2 ( 1 + 2 \beta^2 ) -2 \lambda \kappa \beta^2, \label{eq:62}\\
 & \hat{A} = \left|\lambda A_\lambda - \frac{1}{3} \kappa A_\kappa \beta^2 \right| \beta. \label{eq:63}
\end{align}
\label{eq:64}
\end{subequations}
The extremal value of $H_2$ and the minimum of the potential, 
$V_{B,\mathrm{min}}(\beta)$, along this direction are given 
in the same form as (\ref{eq:10}) and 
(\ref{eq:14}) by replacing $\hat{F}$, $\hat{A}$ and $\hat{m}^2$ 
with (\ref{eq:62}), (\ref{eq:63}) and (\ref{eq:57}). 
Note that the value of $V_{B,\mathrm{min}}(\beta)$ depends on $\beta$.
Then, the following condition is required to stabilize the realistic 
minimum
\begin{align}
 V_{B,\mathrm{min}}(\beta) \ge V_{\mathrm{min}} \label{eq:32},
\end{align}
for any value of $\beta$.

It is important to note here that the case with 
one non-vanishing vev $S$, i.e.,
\begin{align}
H_1 = H_2 =0,~S \neq 0, \label{eq:72}
\end{align}
is included in the direction $B$. When we take the limit $\beta \to \infty$ and $H_2 \to 0$ by keeping $S$ 
to be finite.
If the minimum along this direction 
is the deepest among the ones included in the direction $B$, it can be found automatically by analyzing the 
direction $B$.

%%%%%%%%%%%%%%%%%%%%%%%%%%%%%%%%%%%%%%%
\subsection{$F_S = 0$ and $V_D \neq 0$ direction}\label{sec:f_s=0_and_v_d-neq-0}
%%%%%%%%%%%%%%%%%%%%%%%%%%%%%%%%%%%%%%%
The last direction we analyze is that $F_S = 0$ and $V_D \neq 0$, refered by the direction $C$. 
From $F_S = 0$, we have a relation (\ref{eq:31}) and parametrize the vev's as
\begin{align}
 H_1 = \alpha H_2,\quad S = \sign{S} \beta H_2, \label{eq:65}
\end{align}
where $\alpha$ and $\beta$ should satisfy 
\begin{align}
 \beta^2 = \frac{\lambda}{\kappa} \alpha,\quad (\kappa > 0). \label{eq:66}
\end{align}
The potential is given by (\ref{eq:34}) with 
\begin{subequations}
\begin{align}
 & \hat{F} = \alpha ( 1 + \alpha^2 ) \frac{\lambda^3}{\kappa} + \frac{1}{8} g^2 \left( 1 - \alpha^2 \right)^2, \label{eq:67} \\
 & \hat{A} = \lambda \sqrt{\frac{\lambda}{\kappa}} \left| A_\lambda - \frac{1}{3} A_\kappa \right| \alpha^{3/2}, \label{eq:68} \\
 & \hat{m}^2 = \alpha^2 m_{H_1}^2 + m_{H_2}^2 +  \frac{\lambda}{\kappa} \alpha m_S^2. \label{eq:69}
\end{align}
\label{eq:70}
\end{subequations}

The extremal value of $H_2$ and the minimum of the potential, 
$V_{C,\mathrm{min}}(\alpha)$, are given in the same form as (\ref{eq:10}) and 
(\ref{eq:14}) by replacing $\hat{F}$, $\hat{A}$ and $\hat{m}^2$ with 
(\ref{eq:67}), (\ref{eq:68}) and (\ref{eq:69}).
Note that the value of  $V_{C,\mathrm{min}}(\alpha)$ depends on 
$\alpha$.
If the extremal values satisfy
\begin{align}
 {H_1}_{\mathrm{ext}}^2 + {H_2}_{\mathrm{ext}}^2 = (1+\alpha^2) {H_2}_{\mathrm{ext}}^2 = 
 v^2 \simeq (174~\mathrm{GeV})^2, \label{eq:71}
\end{align}
this minimum can become the true vacuum. Otherwise, it is a false vacuum.
Then, the following condition is required to stabilize the realistic 
minimum
\begin{align}
 V_{C,\mathrm{min}}(\alpha) \ge V_{\mathrm{min}} \label{eq:78},
\end{align}
for any value of $\alpha$.

The direction $C$ includes the direction along which only 
$H_2$ has the non-vanishing vev, i.e.,
\begin{align}
H_1 = S = 0, ~ H_2 \neq 0. \label{eq:77}
\end{align}
This can be easily seen by taking $\alpha = 0$.\footnote{The direction with only $H_1 = 0$ can be found by taking $\alpha \to 0$ 
and $H_2 \to \infty$ keeping $H_1$ finite.} Furthermore the direction $C$ includes the direction $A$ which corresponds to $\alpha = 1$.
Therefore all of the false minima that can be found along the directions proposed so far are included in the direction $B$ and $C$. 
Hence it is enough to analyze the potential along these two directions.

%%%%%%%%%%%%%%
\section{Numerical Analysis}\label{sec:numerical-analysis}
%%%%%%%%%%%%%%
In this section, we present numerical results of the constraints from which the false vacua studied in the previous section 
should not be deeper than the realistic one. In addition, we take into account that (i) physical masses 
of the CP-even, odd and charged Higgs scalars are non-tachyonic, and (ii) the parameters $\lambda,~\kappa$ and the 
top Yukawa coupling have no Landau pole until the GUT scale ($\simeq 1.6 \times 10^{16}$ GeV). For the condition (ii), we 
solve renormalization group (RG) equations at one-loop order \cite{Derendinger:1983bz,Falck:1985aa} from the EWSB scale 
to the GUT scale, requiring that $\lambda$ and $|\kappa|$ are smaller than $2\pi$ at the GUT 
scale \cite{Miller:2003ay,Ellwanger:2009dp}.

%%%%%%%%%%%%%%%%%%
\begin{figure}[t]
\begin{center}
\begin{tabular}{c}
 \includegraphics[height=60mm]{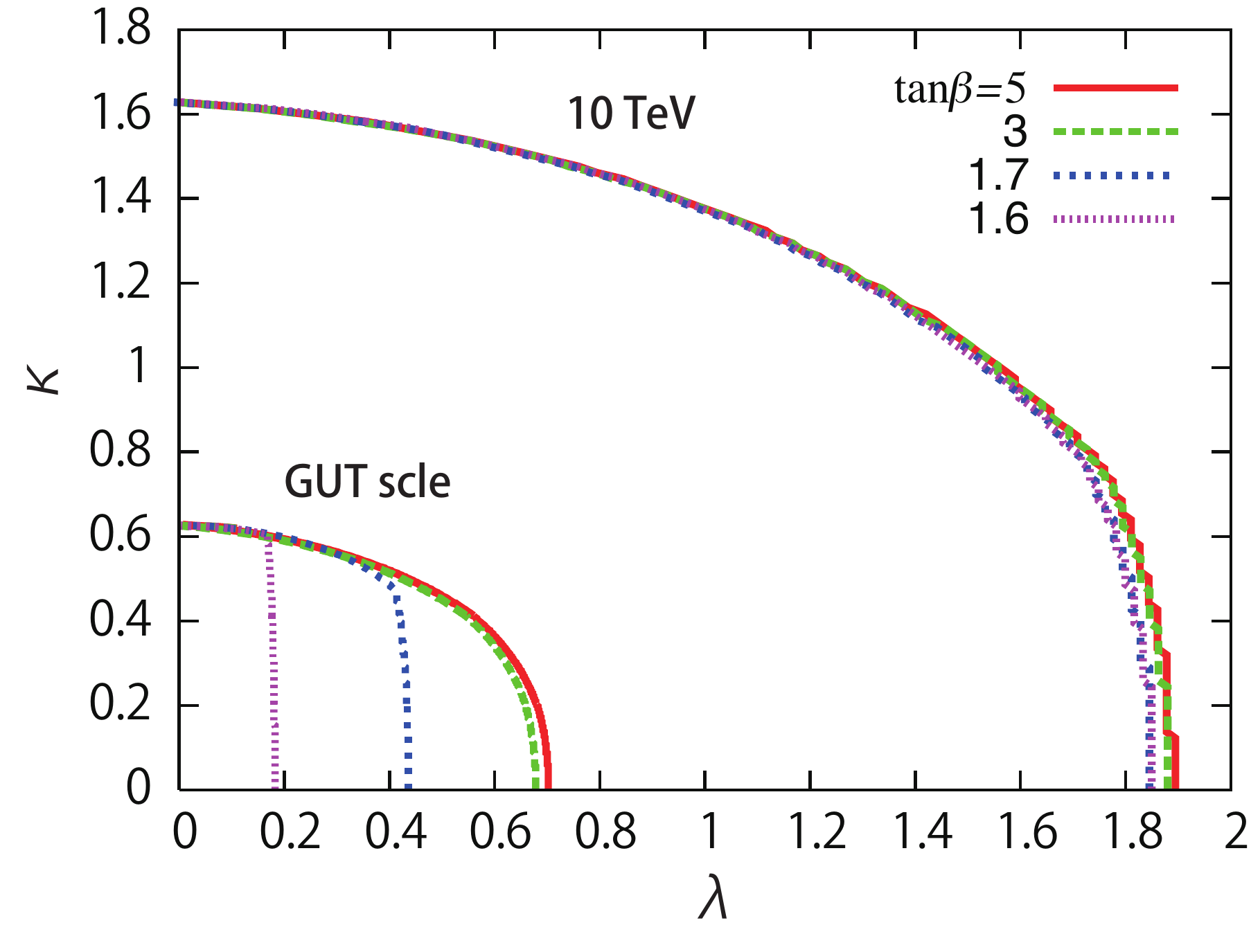}
\end{tabular} 
\caption{Region excluded by the occurrence of a Landau pole on the $\lambda$-$\kappa$ plane. 
Solid (red), dashed (green), dotted (blue) and dashed-dotted (violet) curves 
correspond to $\tan\beta=5,~3,~1.7$ and $1.6$, respectively. The cut-off of the model is taken as the GUT scale (inside) 
and $10$ TeV (outside). The region outside each curve is excluded.}
\label{fig:landau-pole}
\end{center}
\end{figure}
Figure \ref{fig:landau-pole} shows the region excluded by the occurrence of a Landau pole on the $\lambda$-$\kappa$ 
plane. We take a cut-off of the model to be the GUT scale (inside) and $10$ TeV (outside) for references and use the running top quark 
mass, $m_t = 165$ GeV as the input.
Solid (red), dashed (green), dotted (blue) and dashed-dotted (violet) curves 
correspond to $\tan\beta=5,~3,~1.7$ and $1.6$, respectively. The region outside each curve is excluded. 
One can see that, for the GUT scale cut-off, $\lambda$ is more constrained as $\tan\beta$ is smaller while the upper bound on 
$\kappa$ stays constant around $0.63$. This is because RG evolution of 
$\lambda$ is directly connected with the top Yukawa coupling. When $\tan\beta$ is small, 
the top Yukawa coupling at low energy is large and it grows quickly as the energy scale goes up. Then, $\lambda$ 
is driven to a large value as the top Yukawa coupling grows. On the other hand, RG evolution of $\kappa$ is proportional to 
$\kappa^2$ and depends on the top Yukawa coupling only through $\lambda$. Therefore, $\kappa$ starts to grow after 
the top Yukawa coupling and $\lambda$ become sufficiently larger than $2 \pi$. As we can see in figure \ref{fig:landau-pole}, 
the maximum value of $\lambda$ becomes small drastically for $\tan\beta < 2$ and it disappears when $\tan\beta \le 1.5$.
For the $10$ TeV cut-off, the upper bounds on $\lambda$ and $\kappa$ do not change with respect to $\tan\beta$, and are about 
$1.9$ and $1.6$, respectively. The result is understood by the fact that the RG evolutions of $\lambda$ and $\kappa$ are 
determined by the values of these couplings at the EWSB scale, and the evolutions are almost independent of the top 
Yukawa coupling. This is simply because the cut-off is close to the EWSB scale and the top Yukawa coupling 
does not grow very much even for small $\tan\beta$.
In the following, we choose moderate values of $\tan\beta$ to analyze the constraints from the unrealistic minima. 

%%%%%%%
We use the parameter sets given in Table \ref{tab:1} as illustrating examples. 
\begin{table}[t]
\begin{tabular}{|c|c|c|c|c|}\hline
 ~~No.~~ &~~$\tan\beta$~~&~~$\mu$~(GeV)~~&~~$A_\lambda$~(GeV)~~&~~$A_\kappa$~(GeV)~~ \\ \hline\hline
      1        & $3$ & $200$ & $300$  & $-50$ \\ \hline 
      2        & $3$ & $200$ & $-300$ & $-50$ \\ \hline
      3        & $3$ & $400$ & $-300$ & $-50$ \\ \hline
      4        & $3$ & $200$ & $660$  & $-50$ \\ \hline
\end{tabular}
\caption{Parameter sets used in the numerical calculation.}
\label{tab:1}
\end{table}
For radiative corrections to the Higgs potential, we fix the soft masses of the third generation of left-handed squark, 
$m_{\tilde{Q}}^2$, and right-handed stop and sbottom, $m_{\tilde{t}}^2$ and $m_{\tilde{b}}^2$, as 
\begin{align}
 m_{\tilde{Q}_3}^2 = 1000~\mathrm{GeV},~m_{\tilde{t}}^2 = m_{\tilde{b}}^2 = 500~\mathrm{GeV},
\end{align}
respectively. The trilinear term of stop is chosen as nearly maximal mixing so that the lightest Higgs boson mass becomes the largest. 
The point $1$ and $4$ correspond to the case $1$.(a)  and the point $2$ and $3$ correspond to the case $1$.(b) explained 
in Sec.\ref{sec:real-vacu-higgs}, respectively. For the point $4$, $A_\lambda$ is chosen so that the mixing among the doublet and the 
singlet Higgs vanishes for small $\kappa/\lambda$. Note that on these points, the minimum is found for positive $\kappa$.

\begin{figure}[t]
\begin{center}
\begin{tabular}{cc}
 \includegraphics[height=50mm]{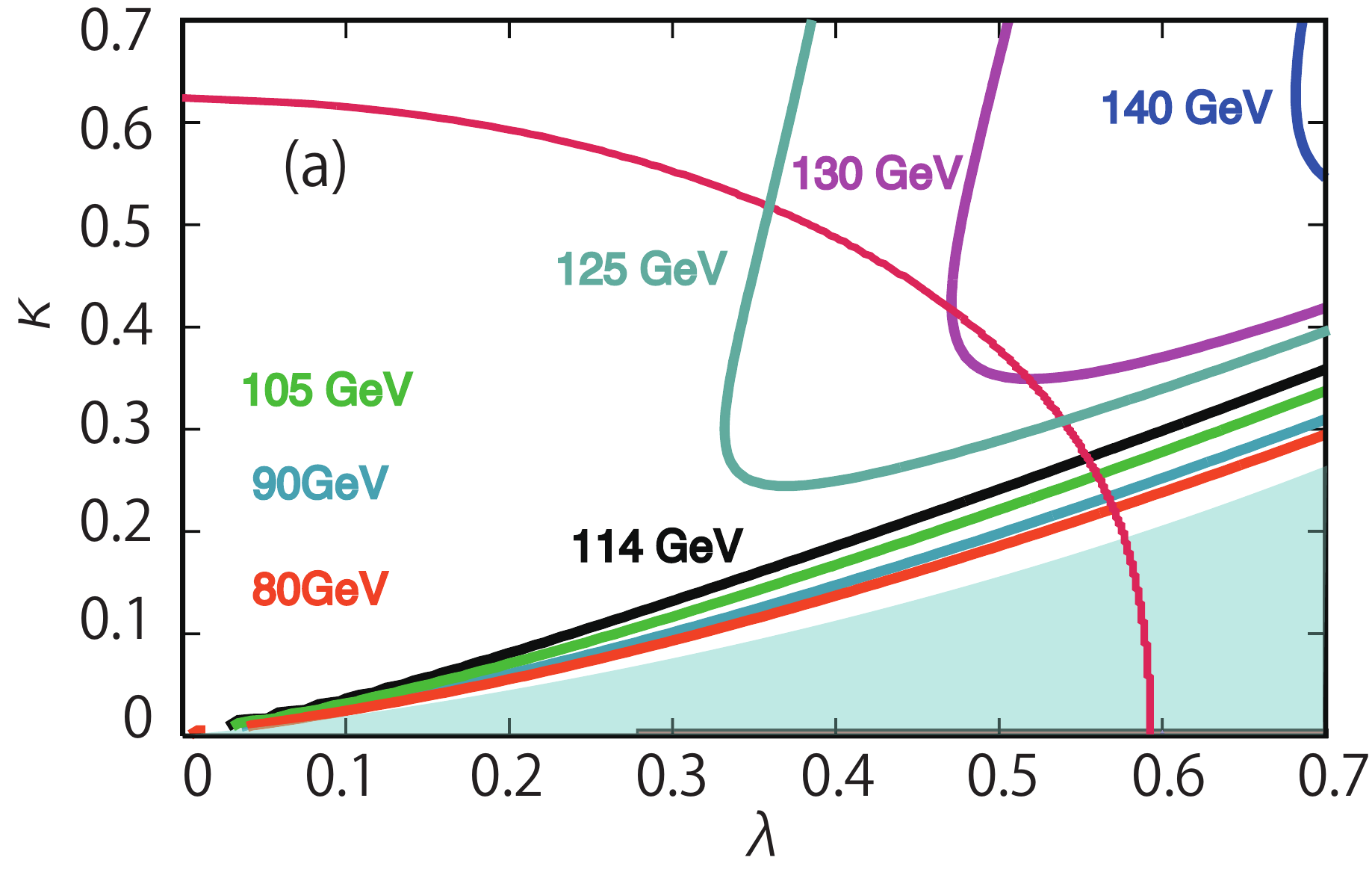} &
 \includegraphics[height=50mm]{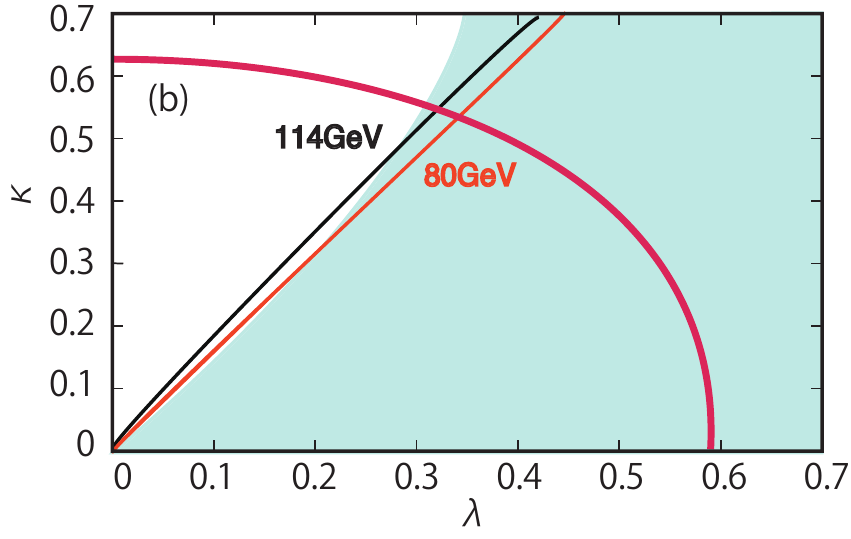} \\
 \includegraphics[height=50mm]{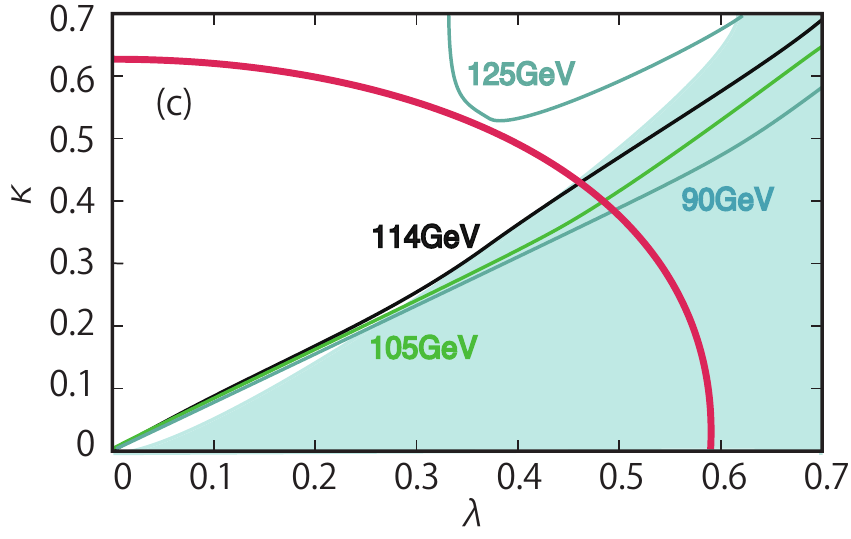} &
 \includegraphics[height=50mm]{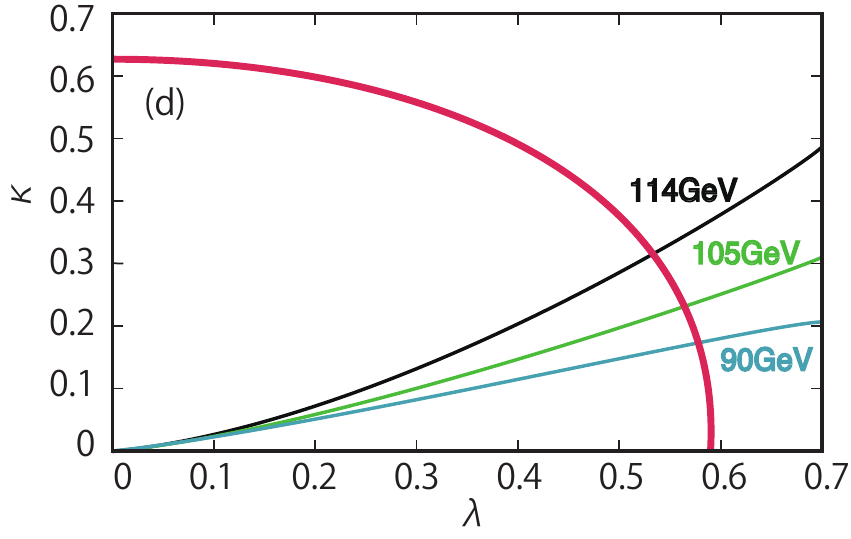}
\end{tabular} 
\caption{Contour plot of the Higgs masses on the $\lambda$-$\kappa$ plane. The values of the Higgs masses are indicated near each curves and left bottom in GeV unit in the figure. The filled region is excluded by the tachyonic Higgs masses and the red solid line represents the Landau 
pole condition.}
\label{fig:tachyonic-higgs}
\end{center}
\end{figure}
In figures \ref{fig:tachyonic-higgs}, we show the contour plot of the lightest CP-even Higgs mass in $\lambda$-$\kappa$ plane. 
The values of the Higgs boson mass are shown near each curve or left bottom in the figure. 
The (light blue) filled region represents the tachyonic mass region of all physical Higgs boson  
and the red solid curve represents the Landau pole condition. 
The figures (a), (b), (c) and (d) correspond to the point $1$, $2$, $3$ and $4$, respectively.
One can see that the tachyonic region appears only in small
$\kappa/\lambda$ region in figure (a) while it also appears 
in large $\kappa/\lambda$ region in figures (b) and (c). The behavior of the tachyonic mass region can be understood 
intuitively by \eqref{eq:30}. Firstly, as we discussed in the Sec.\ref{sec:real-vacu-higgs}, the mass of the CP-even Higgs 
boson tends to become tachyonic if the diagonal elements of the squared-mass matrix is negative. The negative 
diagonal elements can be avoided when \eqref{eq:30} is satisfied for positive $A_\lambda$. 
The second condition of \eqref{eq:30} imposes the lower bound on $\kappa/\lambda \ge |A_\kappa|/4\mu$ for 
negative $A_\kappa$. The lower bound is $0.06$ or $0.03$ for the point $1,~2$ or $3$, respectively. 
Hence the diagonal elements are negative only for very small $\kappa/\lambda$.
Secondly, the Higgs boson mass also tends to become tachyonic when the off-diagonal elements are comparable to the 
diagonal ones. The off-diagonal elements become larger as $\kappa/\lambda$ becomes smaller. Thus the tachyonic 
mass region appears in the small $\kappa/\lambda$ region for positive $A_\lambda$. For negative $A_\lambda$, the first 
condition of \eqref{eq:30} gives a stronger costraint on $\kappa/\lambda$ because the $\mu$ parameter is larger than 
$A_\kappa$. The lower bound on $\kappa/\lambda$ is obtained as $|A_\lambda/\mu| = 1.5$ and $0.75$ for the point $2$ 
and $3$, respectively. Thus, the tachyonic mass region appears for relatively large $\kappa/\lambda$. For the point (d) where the parameters are tuned so that the mixing of the 
doublet and the singlet Higgs bosons vanishes, according to \eqref{eq:29}. In this case, the tachyonic region disappears. 
This is because for small $\kappa/\lambda$, the second term of \eqref{eq:29} is negligible compared to the first term and 
hence the singlet does not mix.  The physical mass of the CP-even Higgs bosons are mainly determined by the diagonal 
elements, and the second condition of \eqref{eq:30} can not be applied in this case.

\begin{figure}[t]
\begin{center}
\begin{tabular}{cc}
 \includegraphics[height=55mm]{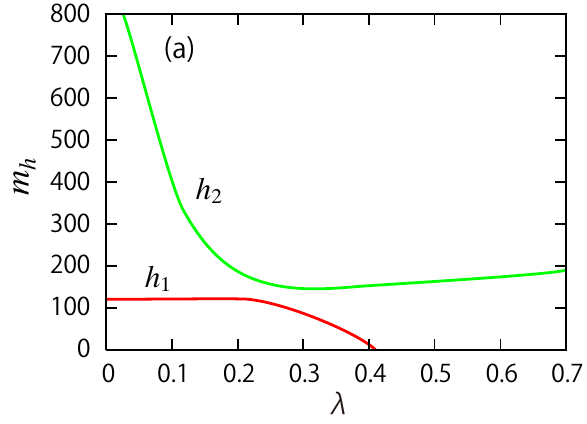} &
 \includegraphics[height=55mm]{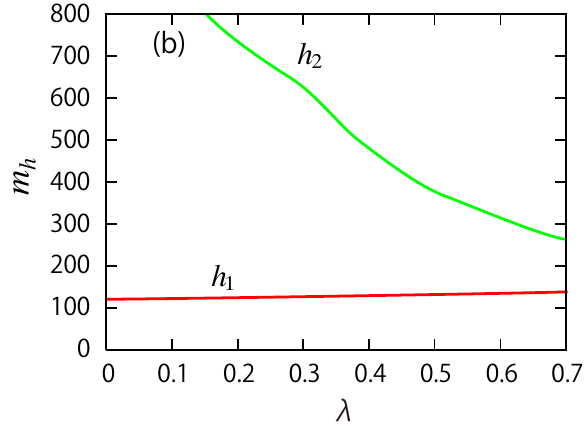}  \\
  \includegraphics[height=55mm]{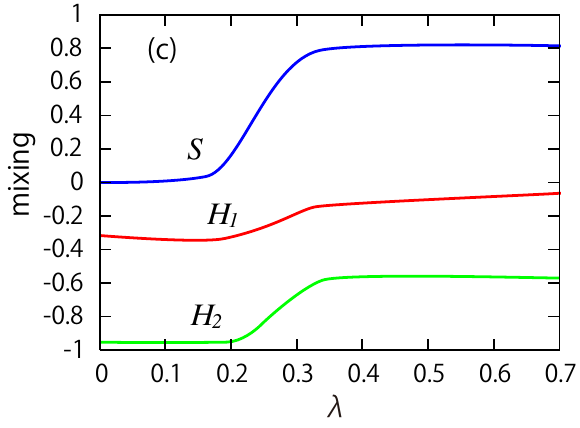} &
  \includegraphics[height=55mm]{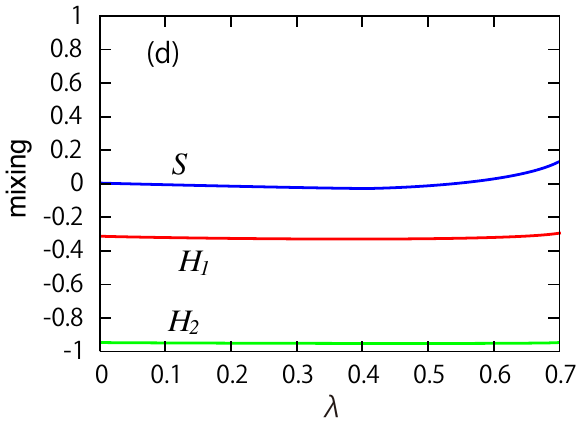}
\end{tabular} 
\caption{The masses of the lightest and the second lightest Higgs boson (top) and the mixings of $H_1$, $H_2$ and $S$ 
in the lightest Higgs boson (bottom) for the point $1$. The left and right figures correspond to $\kappa=0.1$ and $0.5$, respectively.}
\label{fig:h-mass-lambda-dep}
\end{center}
\end{figure}

%%%%%%%%%%%%%%%%%%%%%%%%%%
To see the above explanation more concretely, we show the masses of the Higgs bosons and the mixings in the 
lightest Higgs boson for the point $1$ in Figures \ref{fig:h-mass-lambda-dep}. 
 The mixing, $N_i~(i=1,2,S)$, is defined as 
\begin{align}
 h_1 &= N_1 H_1 + N_2 H_2 + N_S S,  \label{mixing} 
\end{align}
where $h_1$ represents the lightest Higgs boson and $\sum N_i^2 = 1$.
Figures \ref{fig:h-mass-lambda-dep}.(a) and \ref{fig:h-mass-lambda-dep}.(b) show that the masses of the lightest 
Higgs boson $h_1$ and the second lightest Higgs boson $h_2$ for $\kappa=0.1$ and $0.5$, respectively. 
Figures \ref{fig:h-mass-lambda-dep}.(c). and \ref{fig:h-mass-lambda-dep}.(d) show the mixings of the down-type, up-type Higgs 
$H_1$, $H_2$ and the singlet $S$ in the lightest Higgs boson corresponding to (a) and (b).
It is seen in Fig.\ref{fig:h-mass-lambda-dep}.(a) that the mass of $h_1$ is the most degenerate to that of $h_2$ at  
$\lambda = 0.2$ and it becomes tachyonic at $\lambda = 0.4$. From Fig.\ref{fig:h-mass-lambda-dep}.(c), one can see 
that for $\lambda \le 0.2$ the lightest Higgs boson consists of mainly $H_2$ and the mixings of the Higgs bosons 
are constant. The mixings of $H_1$ and $S$ increase at $\lambda = 0.2$ where the mass 
of $h_1$ is the closest to that of $h_2$. The mixings of $H_1$ and $S$ become maximum at $\lambda = 0.4$ where 
the mass of $h_1$ becomes tachyonic. On the other hand,  as is seen in Fig.\ref{fig:h-mass-lambda-dep}.(b), 
the mass of the $h_1$ increases slowly as $\lambda$ increases. The mixing of $H_1$ and $S$ is small and 
the main component of $h_1$ is the up-type Higgs. 
Thus, the tachyonic mass of $h_1$ appears when the mixing of $H_1$ and $S$ becomes sizable.

%%%%%%%%%%%%%%%%
\begin{figure}[t]
\begin{center}
\begin{tabular}{cc}
 \includegraphics[height=57mm]{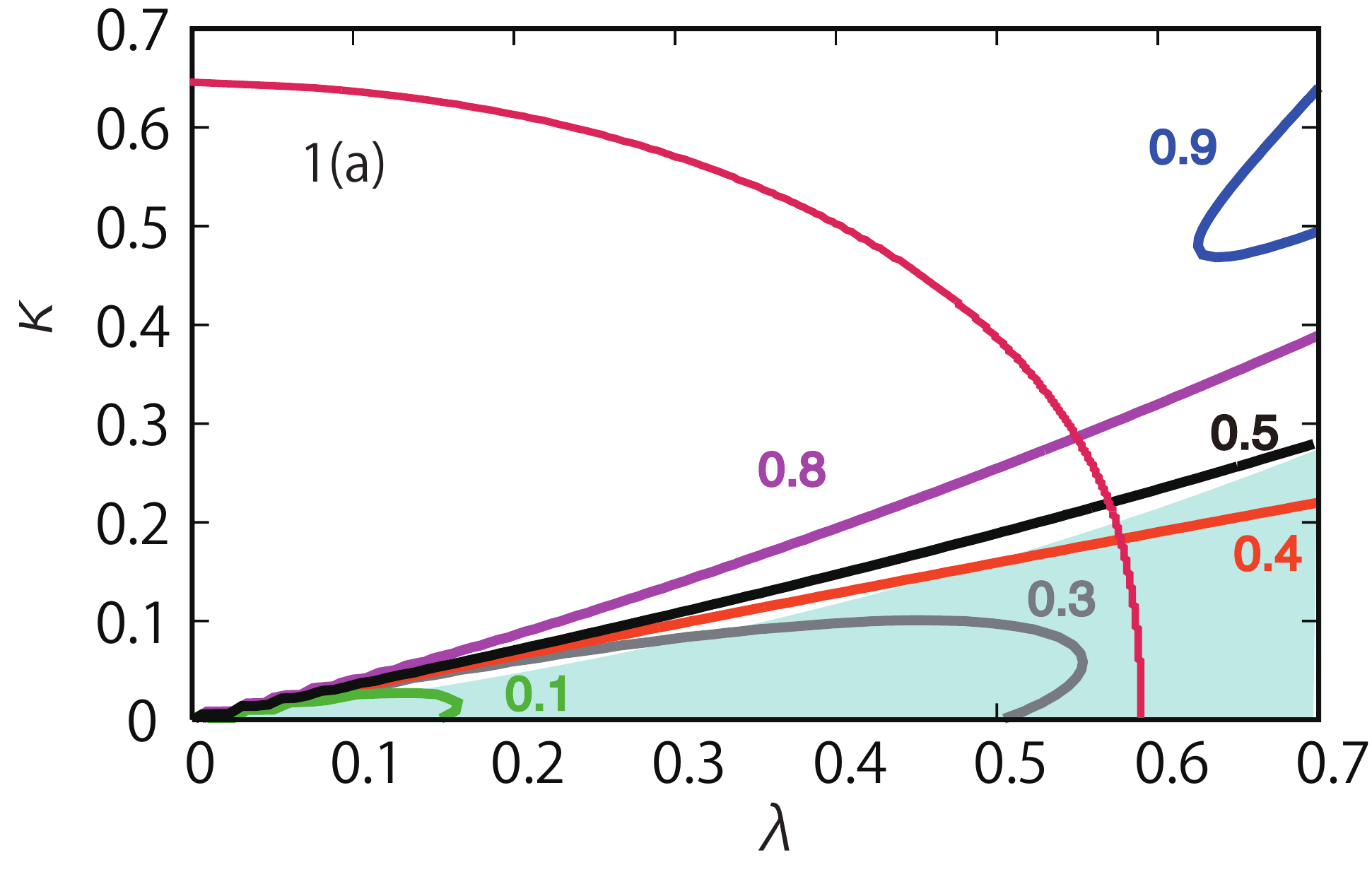} &
 \includegraphics[height=57mm]{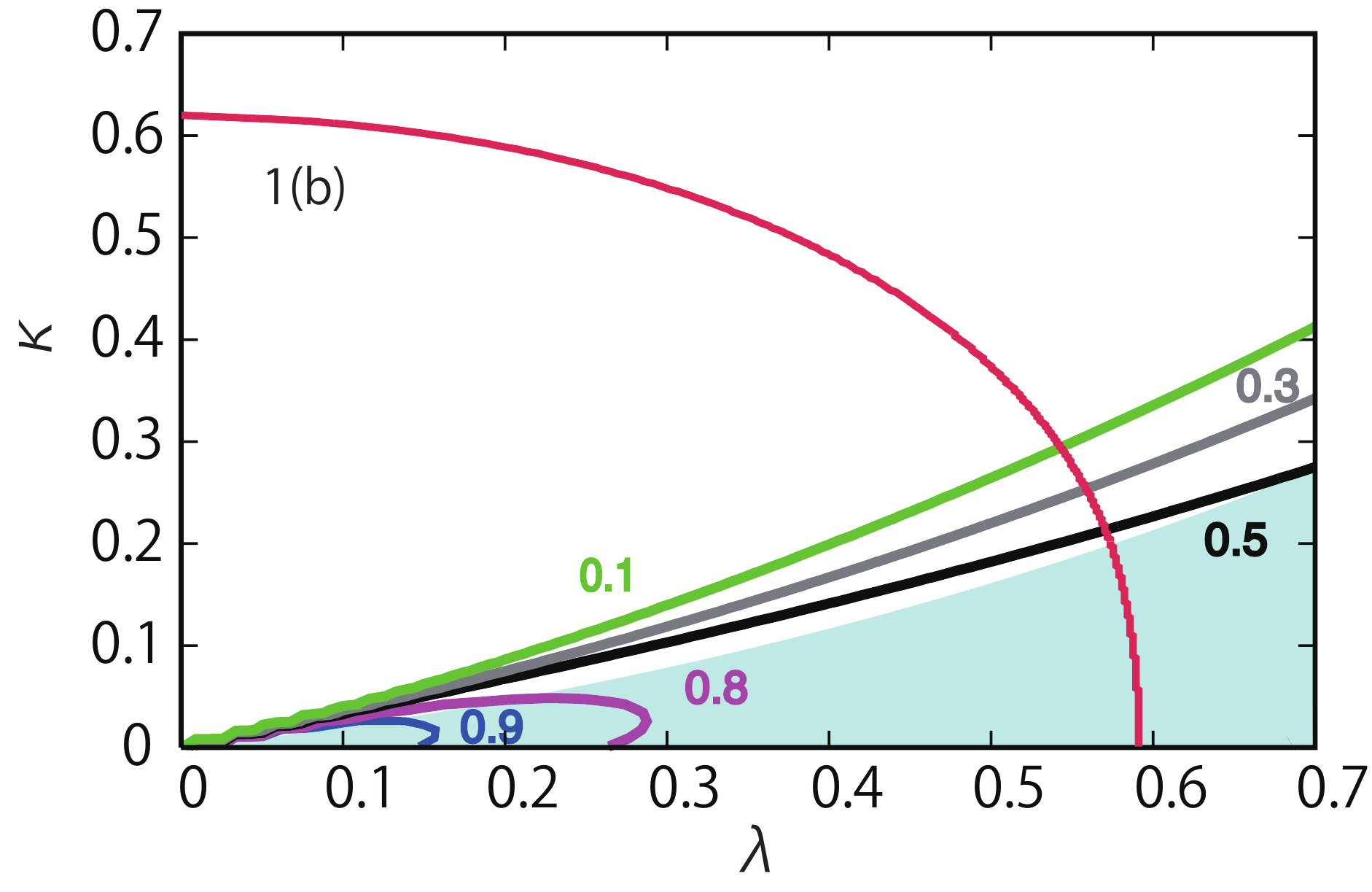}  \\
 \includegraphics[height=55mm]{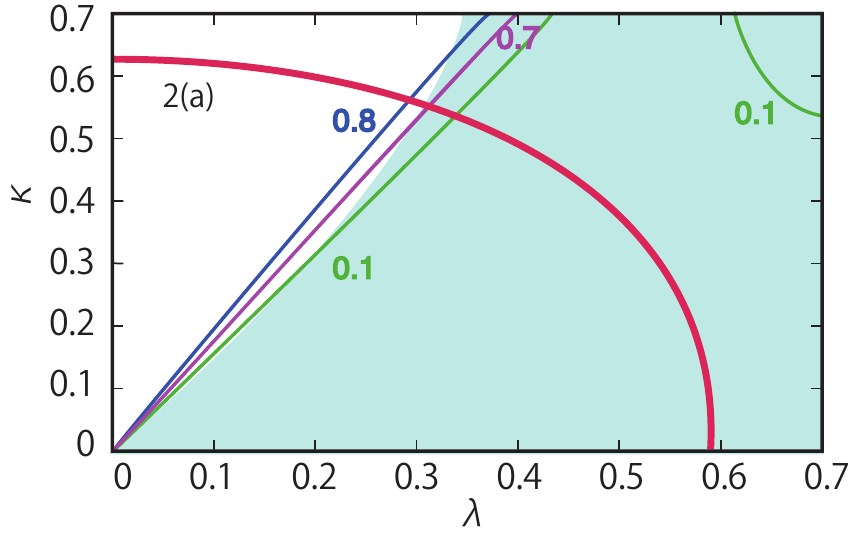} &
 \includegraphics[height=55mm]{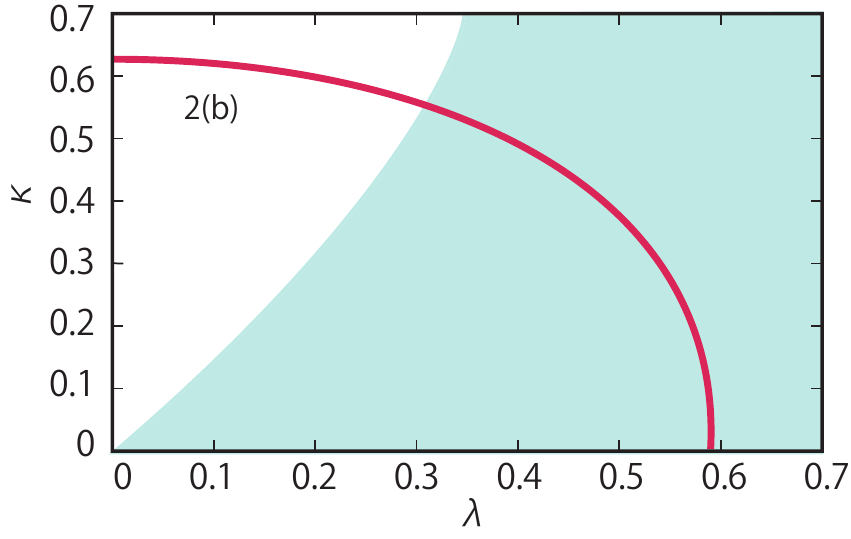}  \\
  \includegraphics[height=55mm]{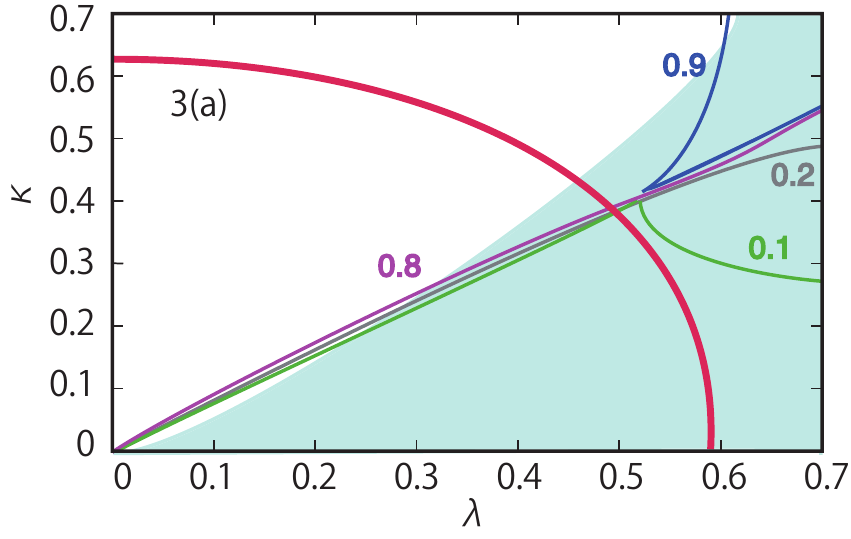} &
 \includegraphics[height=55mm]{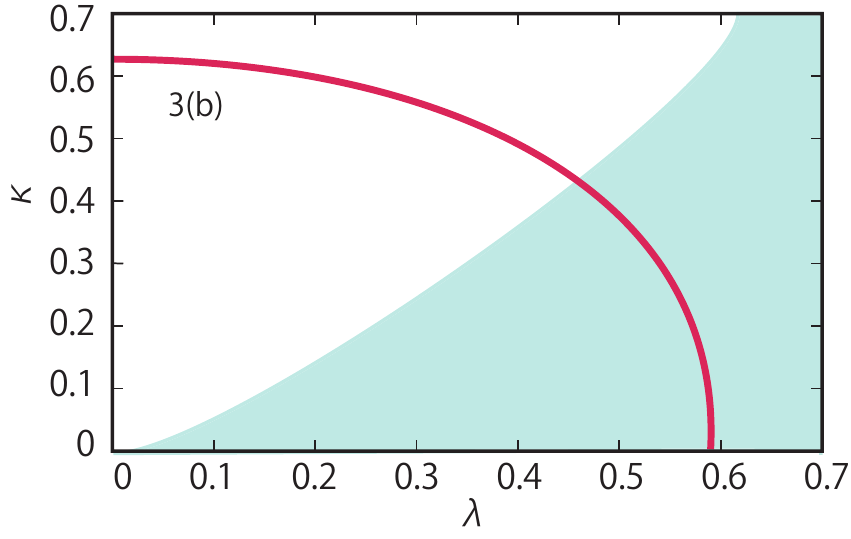} 
\end{tabular} 
\caption{Contour plot of the mixings in the lightest CP-even Higgs boson on the $\lambda$-$\kappa$ plane. }
\label{fig:higgs-mixings}
\end{center}
\end{figure}
Figures \ref{fig:higgs-mixings} show the square of the mixing of the up-type Higgs (a) and the singlet (b) in the lightest CP-even Higgs 
boson on $\lambda$-$\kappa$ plane.
From top to bottom, the figures correspond to the point from $1$ to $3$. The values of the mixing are indicated near each curve 
and the filled region is excluded by the tachyonic Higgses. The red curve represents the Landau pole condition. 
From the Figures 4.1.(a) and 4.1.(b), one can see that  for the point $1$ the tachyonic 
region appears when the squared mixing of $H_2$ ($S$) in the lightest Higgs boson is smaller (larger) than 
about $0.5$. The qualitative behaviour can be understood by the discussion in Sec.~\ref{sec:real-vacu-higgs}. 
On the point $1$, tachyonic region are mainly determined by the CP-even Higgs boson and tachyonic masses of the CP-even 
Higgs boson appear when $\kappa/\lambda$ is small so that (\ref{eq:30}) and (\ref{eq:29}) are not satisfied. 
For small $\kappa/\lambda$, $M_{h,33}^2$ becomes very small and/or because the mixing with the singlet, $M_{h,23}^2$, 
becomes large in this region. The large mixing between the up-type Higgs and the singlet results in large mass splitting between the 
lightest and the second lightest Higgs boson, and hence the mass of the lightest Higgs boson becomes tachyonic.
For the points $2$ and $3$, we can see from Figures $4.2$ and $4.3$ that the tachyonic region appears when the squared mixing of 
the up-type Higgs in the CP-even Higgs boson is smaller than $0.7$ while the mixing of the singlet is almost zero. On these points, 
the tachyonic mass regions are mainly determined by the CP-odd Higgs boson. The diagonal elements of the CP-odd Higgs boson  decreases for the negative $A_\lambda$ while the off-diagonal element increases negatively. Therefore the physical mass of the CP-odd Higgs boson 
becomes tachyonic unless $\kappa/\lambda$ is relatively large. In this situation, the diagonal element
$M_{h,11}^2$ decreases and the off-diagonal element $M_{h,12}^2$ increases for small $\kappa/\lambda$ region due 
to $\tan\beta$ enhancement. Hence the mixing between the up-type and the down-type Higgs bosons significantly reduces 
the lightest Higgs boson mass. The mixing with the singlet does not play important role for this choice of the signs because 
$M_{h,33}^2$ is always larger than $M_{h,11}^2$ even for negative values. Our numerical results show that the up-type 
and the down-type Higgs doublet mix almost maximally in the lightest Higgs boson, the lightest Higgs boson becomes tachyonic.

\begin{figure}[t]
\begin{center}
\begin{tabular}{cc}
 \includegraphics[height=58mm]{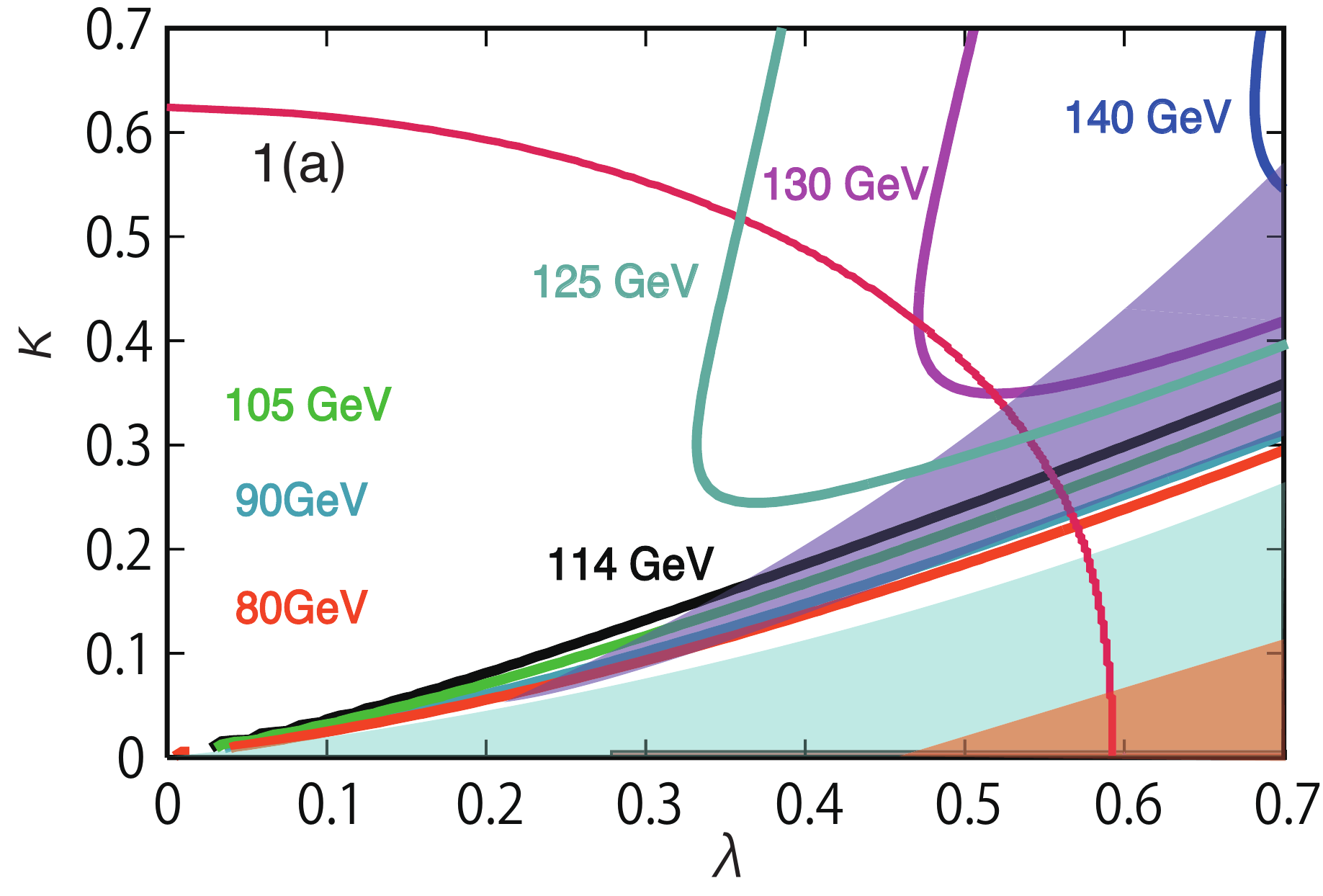} &
 \includegraphics[height=58mm]{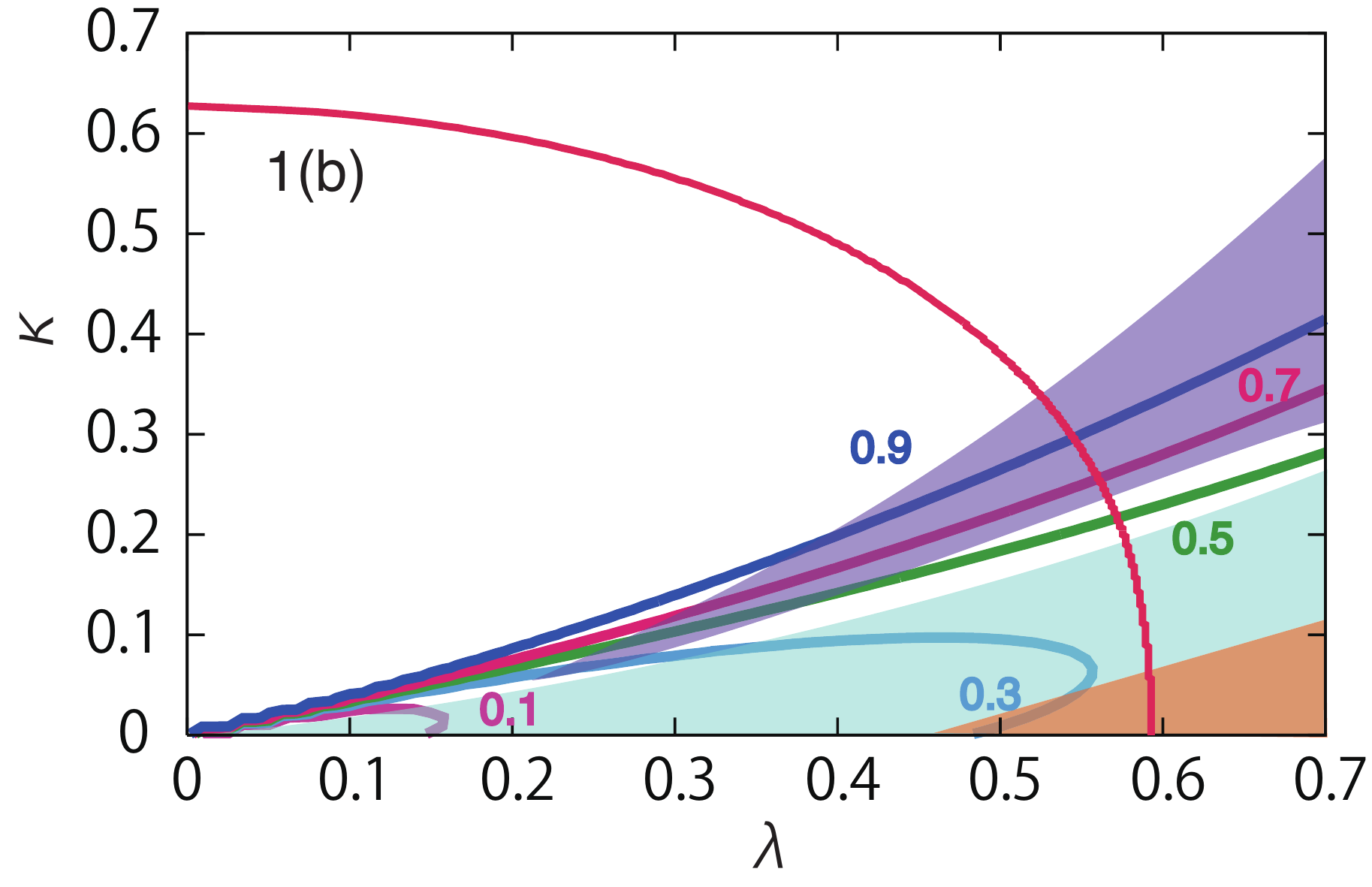} \\
 \includegraphics[height=55mm]{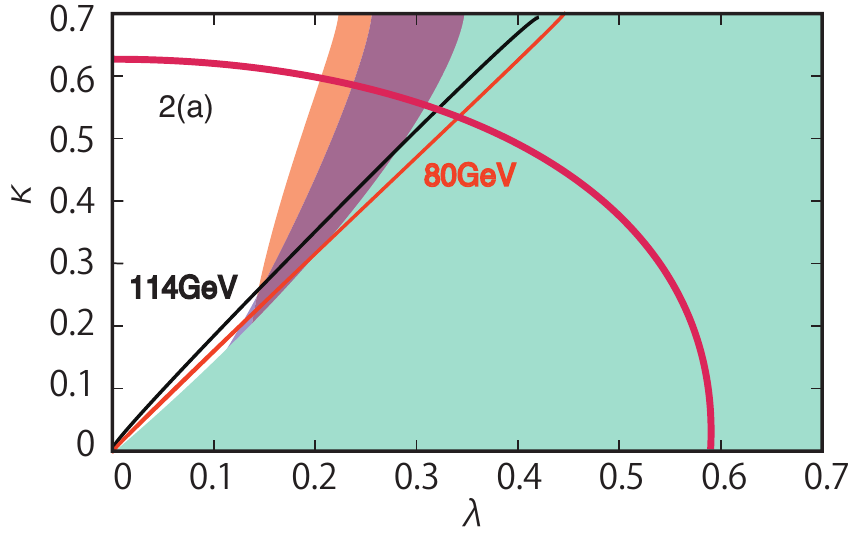} &
 \includegraphics[height=55mm]{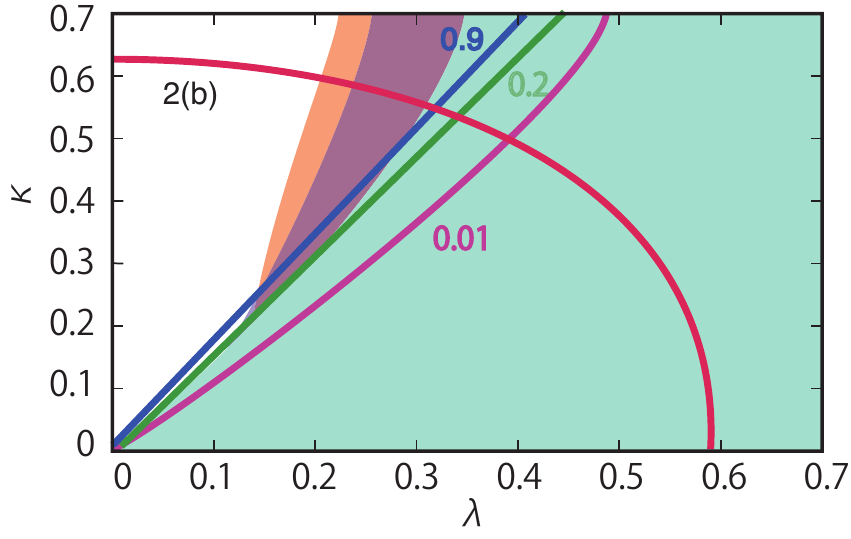} 
\end{tabular} 
\caption{Excluded region of the lightest CP-even Higgs mass (a) and the coupling of the $Z$ boson (b) for the point $1$ and $2$.}
\label{fig:false-vacua1}
\end{center}
\end{figure}

\begin{figure}[t]
\begin{center}
\begin{tabular}{cc}
 \includegraphics[height=55mm]{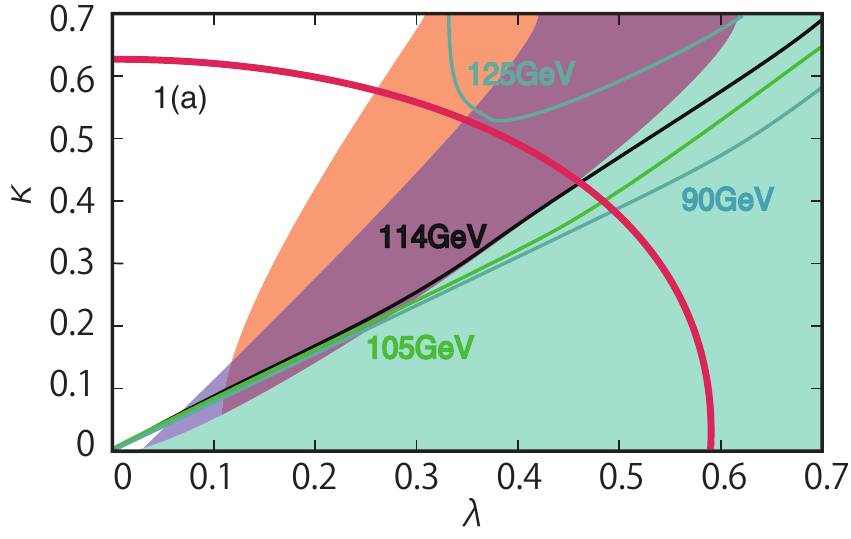} &
 \includegraphics[height=55mm]{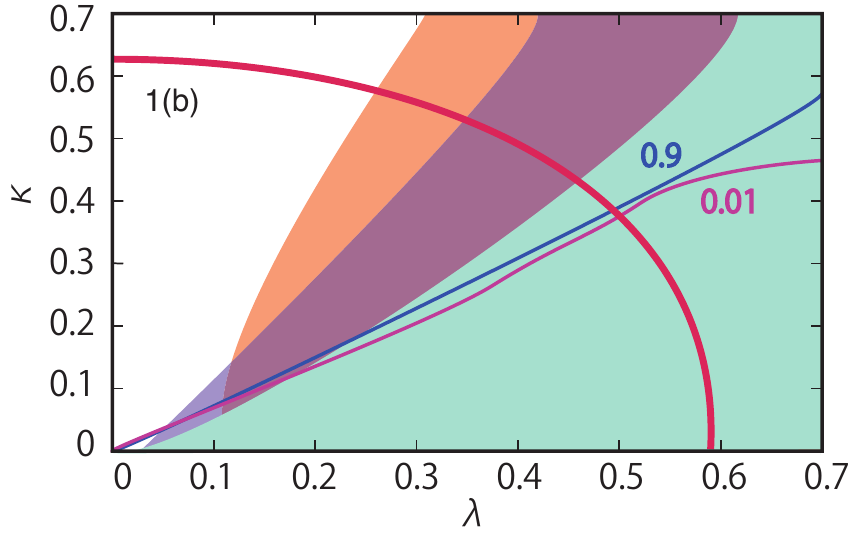} \\
 \includegraphics[height=55mm]{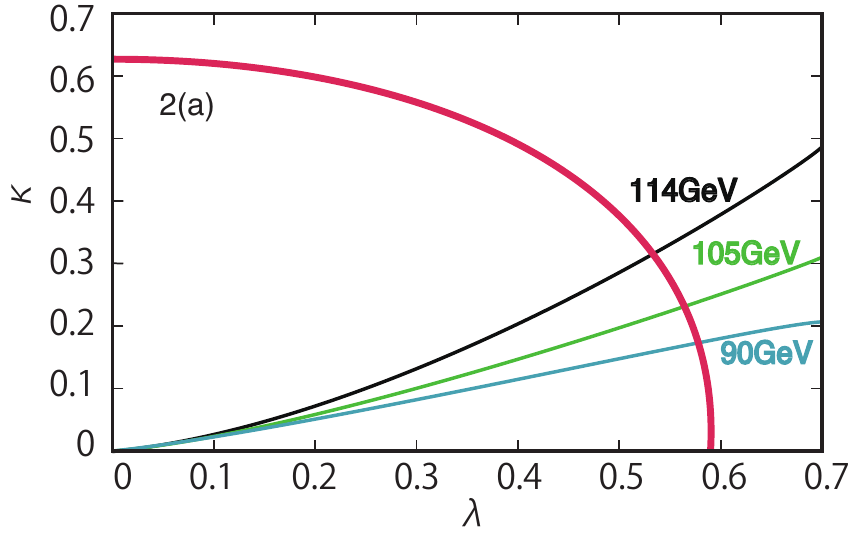} &
 \includegraphics[height=55mm]{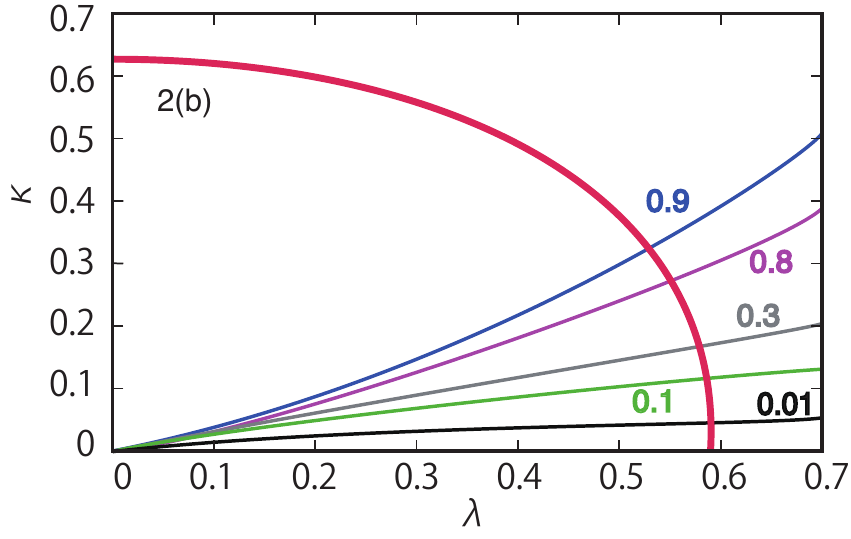}
\end{tabular} 
\caption{The same plots for the point $3$ and $4$ as the figure \ref{fig:false-vacua1}.}
\label{fig:false-vacua2}
\end{center}
\end{figure}

\begin{figure}[t]
\begin{center}
\begin{tabular}{cc}
 \includegraphics[height=55mm]{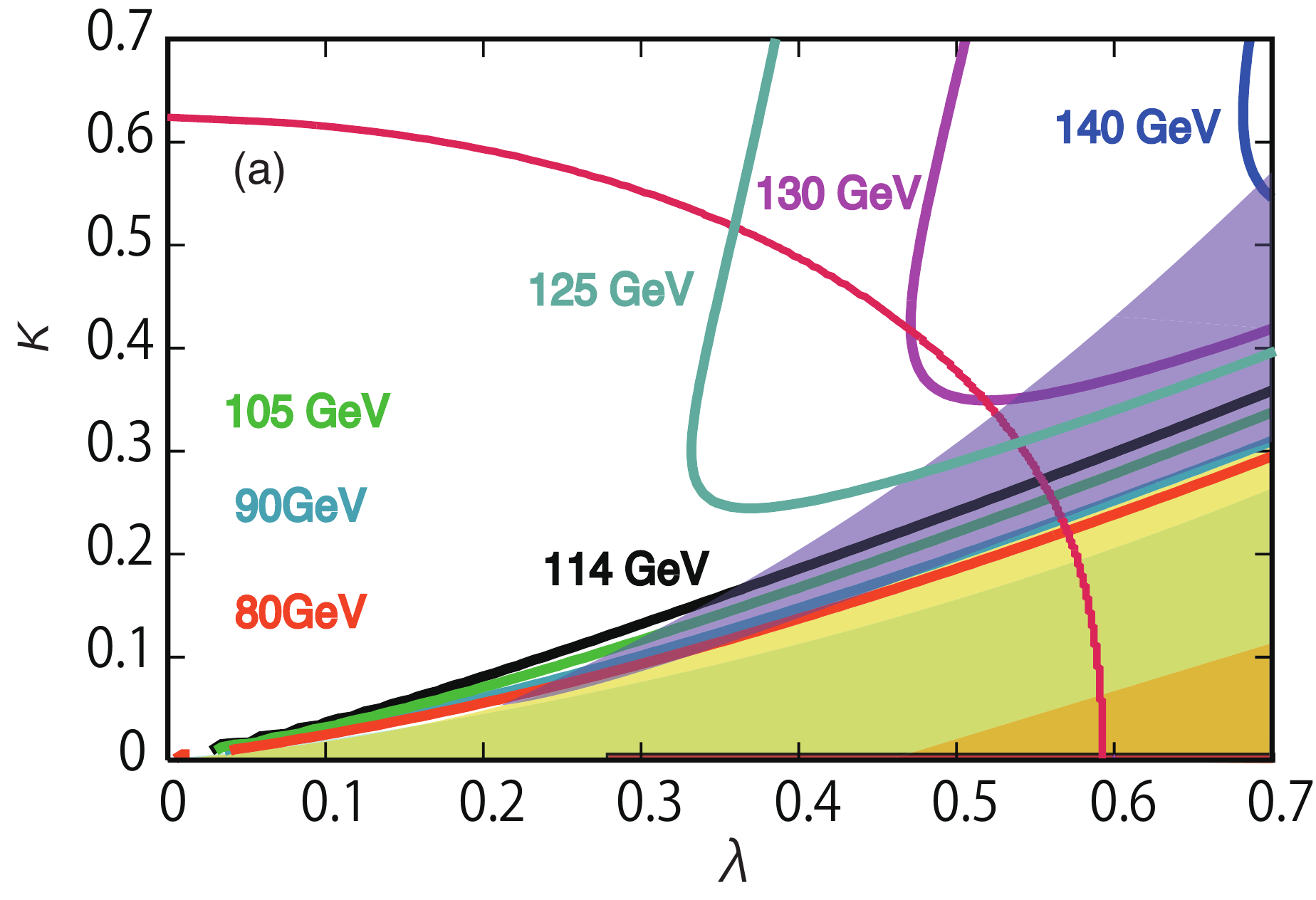} &
 \includegraphics[height=55mm]{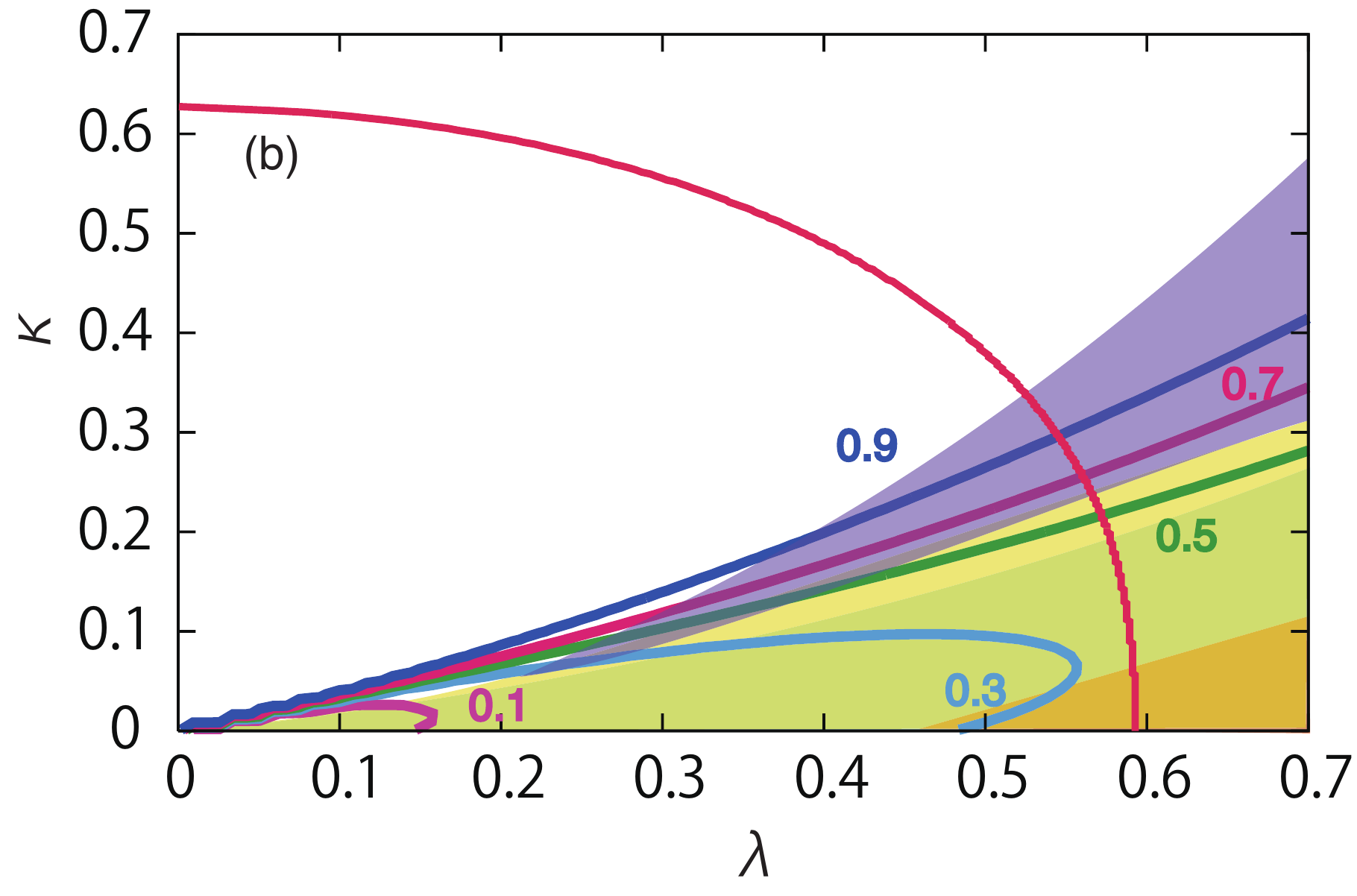} 
\end{tabular} 
\caption{The same figure as fig.\ref{fig:false-vacua1}.1(a) and (b). The region of tachyonic squark masses is indicated by yellow region.}
\label{fig:tachyonic-squarks}
\end{center}
\end{figure}

Figures \ref{fig:false-vacua1} and \ref{fig:false-vacua2} show the constraints from the false vacua on the lightest Higgs mass boson 
(a) and the squared coupling with $Z$ boson normalized by that in the SM (b) in $\lambda$-$\kappa$ plane. 
Figures \ref{fig:false-vacua1}.1 and \ref{fig:false-vacua1}.2 correspond to 
the point $1$ and $2$, and Fig.\ref{fig:false-vacua2}.1 and Fig.\ref{fig:false-vacua2}.2 to the point $3$ and $4$, respectively.
As the same as in Fig.~\ref{fig:tachyonic-higgs}, the values of the lightest Higgs boson mass are indicated 
near the curves in Figures (a) and those of the squared coupling with $Z$ boson are indicated in figures (b). 
The regions excluded by the constraints from the direction $A$, $B$ and $C$ are indicated by filled region coloured 
as green, orange and purple, respectively. The blue filled region represents tachyonic region. 
In figure \ref{fig:false-vacua1}.1(a), one can see that the constraint from the direction $C$ excludes the region 
outside the tachyonic region while the constraints from the direction $B$ excludes inside the tachyonic region. 
The region excluded by the direction $A$ is very narrow and appears near the horizontal axis. The region excluded by 
the direction $B$ is simply connected from the realistic vaccum along the tachyonic region. However, the region excluded 
by the direction $C$ appears outside the tachyonic region, and hence is not connected to the realistic minimum. 
The region can not be found by only taking tachyonic masses into account. 
From the figure \ref{fig:false-vacua1}.1(a), one can see that it excludes the large region of the lightest Higgs boson mass. 
From the \ref{fig:false-vacua1}.1(b), one can see that the constraint from the $C$ direction excludes the region of the squared 
coupling from $0.7$ to $0.9$. This implies that the lightest Higgs boson has the SM-like coupling with $Z$ boson and 
consists of almost purely the up-type Higgs in most of allowed region. It is noted that there exists small region 
allowed between the tachyonic mass region and the region excluded by the direction $C$. In this region, the coupling with $Z$ 
boson is about half of that in the SM. However, as shown in figure \ref{fig:tachyonic-squarks}, squard masses 
become tachyonic in this region. Thus, this region is not realistic because colour and charge symmetry are spontaneously broken.
In the figures \ref{fig:false-vacua1}.2(a) and \ref{fig:false-vacua2}.1(a), the constraints from the direction $B$ and $C$ exclude regions outside the tachyonic region. 
The constraint from the direction $A$ also excludes the similar but smaller regions. From these figures, large regions on 
$\lambda$-$\kappa$ plane are excluded. The allowed region corresponds to small $\lambda$ region where the mass of the 
lightest Higgs boson is smaller than $120$ GeV and the squared coupling is the SM-like.  
For the point $4$, we can see from the figure
\ref{fig:false-vacua2}.2(a) and \ref{fig:false-vacua2}.2(b) that there
is no region excluded by the false vacua. As explained before, the
mixing of the Higgs bosons is vanishing due to the choice of parameters satisfying \eqref{eq:29}. 
In this case, the false vacua along the direction $A$, $B$ and $C$ are not deeper than the realistic vacuum, and hence the 
realistic minimum is stable. The lightest Higgs boson mass as well as the squared coupling with $Z$ boson are not constrained at all. However, an unnatural tuning between $A_\lambda$, $\mu$ and $\tan\beta$ is required. 
The behavior of the constraints or the depth of the false minima on the parameters is very complicated and it is difficult 
to understand qualitatively. However, as we have seen, our constraint can exclude sizable region on the parameter 
space which can not be found by tachyonic Higgs mass. Thus, it is important to include the constraints from the false vacua 
in phenomenological studies.

\begin{figure}[t]
\begin{center}
\begin{tabular}{cc}
 \includegraphics[height=55mm]{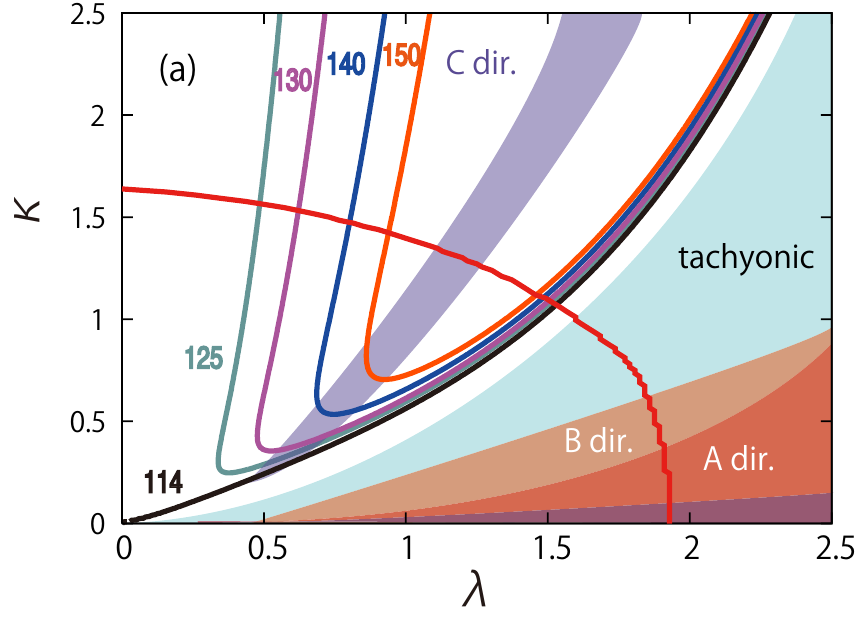} &
  \includegraphics[height=55mm]{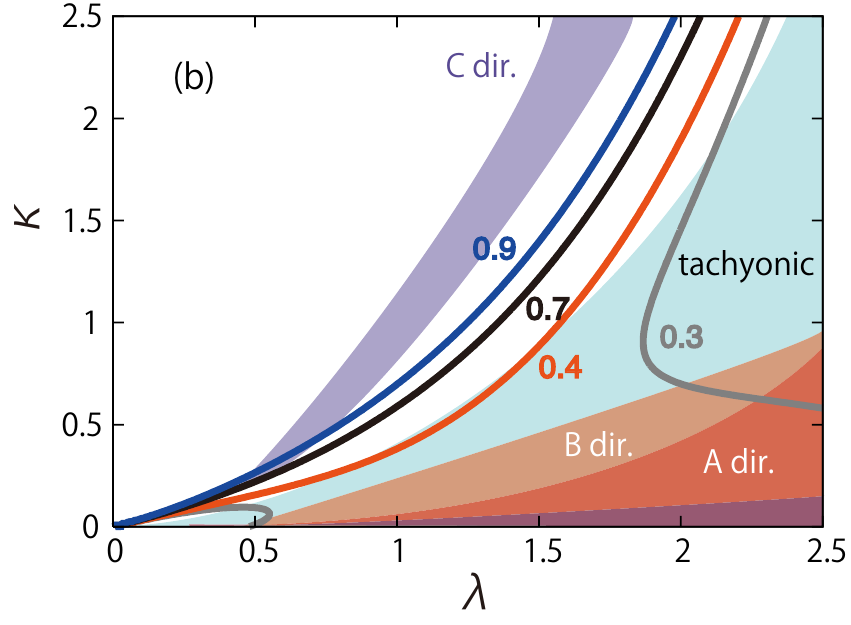} 
\end{tabular} 
\caption{The same figure as figure \ref{fig:false-vacua1}.1(a) and 1(b) for the cut-off $10$ TeV. }
\label{fig:wide-point1}
\end{center}
\end{figure}

So far, we have shown numerical resuls for the GUT scale cut-off. The NMSSM with the $10$ TeV cut-off scale is also interesting 
in the view of the heavy Higgs boson without little hierarchy \cite{Barbieri:2006bg,Franceschini:2010qz,Lodone:2011ax,Bertuzzo:2011aa}. We show the constraints for the point $1$ with 
$10$ TeV cut-off in figure \ref{fig:wide-point1}. One can see that the large region on $\lambda$-$\kappa$ plane is excluded by our constraints although the lightest Higgs boson with the mass $125$ GeV and the SM-like coupling can be obtained. In the region between 
the tachyonic Higgs boson mass and the excluded region by the C direction, squark masses become tachyonic and hence the region is 
not allowed phenomenologically.

%%%%%%%%%%%%%%%%%%%%
\section{Conclusion}\label{sec:conclusion}
%%%%%%%%%%%%%%%%%%%%
We have studied the mass, the mixings and the couplings with $Z$ boson of the lightest Higgs boson 
in the viewpoint of the structure of vacua in the Next-to-Minimal Supersymmetric Standard Model. 

In section \ref{sec:real-vacu-higgs}, we have shown the intuitive  
conditions for which the tachyonic masses for the CP-even and odd Higgs bosons can be avoided. The conditions are 
derived by requiring that the diagonal elements should not be negative and the off-diagonal elements are not comparable 
to the diagonal ones. These give rough bounds on the parameters which are useful to understand the behavior of the tachyonic masses.
In section \ref{sec:false-vacua-along}, we have shown the new directions, along which unrealistic vacua can appear. 
The EW symmetry is not broken successfully on these unrealistic vacua and hence these vacua should not be chosen 
as our vacuum. We have seen that the depth of the false vacua is characterized by the SUSY breaking scale. 
Therefore the false vacua can become deeper than the realistic vacuum. We have also shown that the false vacua studied in the previous 
works \cite{Kanehata:2011ei} are included in the new directions. 

In section \ref{sec:numerical-analysis}, we have shown our numerical results on the mass. 
First we have seen that the region of tachyonic Higgs mass appears for small $\kappa/\lambda$ for positive $A_\lambda$ and 
negative $A_\kappa$ while it also appears for large $\kappa/\lambda$ for negative $A_\lambda$ and $A_\kappa$. 
In the former case, the lightest CP-even Higgs boson becomes tachyonic when the mixing of the singlet becomes larger than $0.5$, 
while in the latter case, tachyonic mass of the CP-odd Higgs boson appears when the mixing of the up-type Higgs becomes larger than 
$0.7$.  Then, we have shown that the new false vacua appear outside the tachyonic Higgs mass region and 
large $\kappa/\lambda$ region. This result implies that the mass and the mixing of the lightest Higgs boson can not be large because 
large value of $\lambda$ is excluded in both cases. 
In fact, we have shown that by imposing the constraint that the realistic vacuum is deeper 
than the new false vacua, important parameter regions for the Higgs mass around $125$ GeV can be excluded. 
The large mixing of the up-tye, down-type and the singlet Higgs in the lightest Higgs boson is also excluded by the new false vacua, 
and the lightest Higgs boson consists of mainly the up-type Higgs boson in the allowed region.
Then, we have seen that the squared coupling with $Z$ boson of the lightest Higgs boson is very close to that of the SM.
On the other hand, we have seen that the mass and the mixing do not constrain by the tachyonic Higgs mass and the false vacua 
if the parameters satisfy \eqref{eq:29}. In this case, however, the parameters should be tuned so that the mixing of the 
doublet and the singlet Higgs boson vanishes.

In general, our constraints exclude wider parameter region than 
the constraint to avoid the tachyonic Higgs masses.
In most of cases, the region with small $\kappa/\lambda$ 
is excluded.
Furthermore, the component of the up-type Higgs in 
the lightest Higgs boson is also constrained.
In most of cases, the region with such component less than 
$0.9$ is disfavored.
That implies that the lightest Higgs boson is SM-like.
However, the region with (\ref{eq:29}) is exceptional, 
the tachyonic modes do not appear and the false vacua 
are less deep than the realistic vacuum.

In the end of the conclusion, we comment on the lifetime of the false vacua. As was discussed in \cite{Kanehata:2011ei}, 
if the lifetime of vacua is longer than the age of the universe, a realistic vacuum becomes metastable and 
the parameter space is not constrained. A Euclidean action for bounce solutions \cite{Coleman:1977py,Coleman:1977py2,Callan:1977pt,Coleman:1980aw,Lee:1985uv,Duncan:1992ai} should be larger than $400$ 
for the lifetime to be longer than the age of the universe. We estimated the Euclidian action for the new false 
vacuum and found the order of $10$ - $100$. Thus, our results will be still valid if the lifetime is 
taken into consideration. However, detailed studies of the lifetime are important to obtain more serious constraints. 
We will study these aspects in our future works. At any rate, the NMSSM has several interesting aspects.
It is important to study those aspects of the NMSSM by taking into account our new false vacua.

%%%%%%%%%%
\acknowledgments
%%%%%%%%%%
The authors would like to thank Tevon You for fruitful discussions.
T.~K. is supported in part by a Grant-in-Aid for Scientific Research No.~20540266 and the Grant-in-Aid for the 
Global COE Program ``The Next Generation of Physics, Spun from Universality and Emergence'' from the Ministry of 
Education, Culture, Sports, Science and Technology of Japan. 
T.~S. is a Yukawa Fellow and his work is partially supported by the Yukawa Memorial Foundation, 
a Graint-in-Aid for Young Scientists (B) No.~23740190 and the Sasakawa Scientific Research Grant from The Japan Science Society.

%%%%%%%%%%%%%%%%%%
%%% references %%%
%%%%%%%%%%%%%%%%%%
\bibliographystyle{apsrev}
\bibliography{biblio}

\begin{thebibliography}{50}
\expandafter\ifx\csname natexlab\endcsname\relax\def\natexlab#1{#1}\fi
\expandafter\ifx\csname bibnamefont\endcsname\relax
  \def\bibnamefont#1{#1}\fi
\expandafter\ifx\csname bibfnamefont\endcsname\relax
  \def\bibfnamefont#1{#1}\fi
\expandafter\ifx\csname citenamefont\endcsname\relax
  \def\citenamefont#1{#1}\fi
\expandafter\ifx\csname url\endcsname\relax
  \def\url#1{\texttt{#1}}\fi
\expandafter\ifx\csname urlprefix\endcsname\relax\def\urlprefix{URL }\fi
\providecommand{\bibinfo}[2]{#2}
\providecommand{\eprint}[2][]{\url{#2}}

\bibitem[{\citenamefont{Kim and Nilles}(1984)}]{Kim:1983dt}
\bibinfo{author}{\bibfnamefont{J.~E.} \bibnamefont{Kim}} \bibnamefont{and}
  \bibinfo{author}{\bibfnamefont{H.~P.} \bibnamefont{Nilles}},
  \bibinfo{journal}{Phys. Lett.} \textbf{\bibinfo{volume}{B138}},
  \bibinfo{pages}{150} (\bibinfo{year}{1984}).

\bibitem[{1088223()}]{:2012si}
1088223 (\bibinfo{year}{2012}), \eprint{1202.1408}.

\bibitem[{\citenamefont{Chatrchyan et~al.}(2012)}]{Chatrchyan:2012tx}
\bibinfo{author}{\bibfnamefont{S.}~\bibnamefont{Chatrchyan}}
  \bibnamefont{et~al.} (\bibinfo{collaboration}{CMS Collaboration})
  (\bibinfo{year}{2012}), \eprint{1202.1488}.

\bibitem[{\citenamefont{Fayet}(1975)}]{Fayet:1974pd}
\bibinfo{author}{\bibfnamefont{P.}~\bibnamefont{Fayet}},
  \bibinfo{journal}{Nucl. Phys.} \textbf{\bibinfo{volume}{B90}},
  \bibinfo{pages}{104} (\bibinfo{year}{1975}).

\bibitem[{\citenamefont{Fayet}(1976)}]{Fayet:1976et}
\bibinfo{author}{\bibfnamefont{P.}~\bibnamefont{Fayet}},
  \bibinfo{journal}{Phys. Lett.} \textbf{\bibinfo{volume}{B64}},
  \bibinfo{pages}{159} (\bibinfo{year}{1976}).

\bibitem[{\citenamefont{Fayet}(1977)}]{Fayet:1977yc}
\bibinfo{author}{\bibfnamefont{P.}~\bibnamefont{Fayet}},
  \bibinfo{journal}{Phys. Lett.} \textbf{\bibinfo{volume}{B69}},
  \bibinfo{pages}{489} (\bibinfo{year}{1977}).

\bibitem[{\citenamefont{Fayet}(1979)}]{Fayet:1979sa}
\bibinfo{author}{\bibfnamefont{P.}~\bibnamefont{Fayet}},
  \bibinfo{journal}{Phys. Lett.} \textbf{\bibinfo{volume}{B84}},
  \bibinfo{pages}{416} (\bibinfo{year}{1979}).

\bibitem[{\citenamefont{Nilles et~al.}(1983)\citenamefont{Nilles, Srednicki,
  and Wyler}}]{Nilles:1982dy}
\bibinfo{author}{\bibfnamefont{H.~P.} \bibnamefont{Nilles}},
  \bibinfo{author}{\bibfnamefont{M.}~\bibnamefont{Srednicki}},
  \bibnamefont{and} \bibinfo{author}{\bibfnamefont{D.}~\bibnamefont{Wyler}},
  \bibinfo{journal}{Phys. Lett.} \textbf{\bibinfo{volume}{B120}},
  \bibinfo{pages}{346} (\bibinfo{year}{1983}).

\bibitem[{\citenamefont{Frere et~al.}(1983)\citenamefont{Frere, Jones, and
  Raby}}]{Frere:1983ag}
\bibinfo{author}{\bibfnamefont{J.~M.} \bibnamefont{Frere}},
  \bibinfo{author}{\bibfnamefont{D.~R.~T.} \bibnamefont{Jones}},
  \bibnamefont{and} \bibinfo{author}{\bibfnamefont{S.}~\bibnamefont{Raby}},
  \bibinfo{journal}{Nucl. Phys.} \textbf{\bibinfo{volume}{B222}},
  \bibinfo{pages}{11} (\bibinfo{year}{1983}).

\bibitem[{\citenamefont{Derendinger and Savoy}(1984)}]{Derendinger:1983bz}
\bibinfo{author}{\bibfnamefont{J.~P.} \bibnamefont{Derendinger}}
  \bibnamefont{and} \bibinfo{author}{\bibfnamefont{C.~A.} \bibnamefont{Savoy}},
  \bibinfo{journal}{Nucl. Phys.} \textbf{\bibinfo{volume}{B237}},
  \bibinfo{pages}{307} (\bibinfo{year}{1984}).

\bibitem[{\citenamefont{Ellis et~al.}(1989)\citenamefont{Ellis, Gunion, Haber,
  Roszkowski, and Zwirner}}]{Ellis:1988er}
\bibinfo{author}{\bibfnamefont{J.~R.} \bibnamefont{Ellis}},
  \bibinfo{author}{\bibfnamefont{J.~F.} \bibnamefont{Gunion}},
  \bibinfo{author}{\bibfnamefont{H.~E.} \bibnamefont{Haber}},
  \bibinfo{author}{\bibfnamefont{L.}~\bibnamefont{Roszkowski}},
  \bibnamefont{and} \bibinfo{author}{\bibfnamefont{F.}~\bibnamefont{Zwirner}},
  \bibinfo{journal}{Phys. Rev.} \textbf{\bibinfo{volume}{D39}},
  \bibinfo{pages}{844} (\bibinfo{year}{1989}).

\bibitem[{\citenamefont{Drees}(1989)}]{Drees:1988fc}
\bibinfo{author}{\bibfnamefont{M.}~\bibnamefont{Drees}}, \bibinfo{journal}{Int.
  J. Mod. Phys.} \textbf{\bibinfo{volume}{A4}}, \bibinfo{pages}{3635}
  (\bibinfo{year}{1989}).

\bibitem[{\citenamefont{Ellwanger et~al.}(2010)\citenamefont{Ellwanger,
  Hugonie, and Teixeira}}]{Ellwanger:2009dp}
\bibinfo{author}{\bibfnamefont{U.}~\bibnamefont{Ellwanger}},
  \bibinfo{author}{\bibfnamefont{C.}~\bibnamefont{Hugonie}}, \bibnamefont{and}
  \bibinfo{author}{\bibfnamefont{A.~M.} \bibnamefont{Teixeira}},
  \bibinfo{journal}{Phys. Rept.} \textbf{\bibinfo{volume}{496}},
  \bibinfo{pages}{1} (\bibinfo{year}{2010}), \eprint{0910.1785}.

\bibitem[{\citenamefont{Ellwanger}(2011)}]{Ellwanger:2011sk}
\bibinfo{author}{\bibfnamefont{U.}~\bibnamefont{Ellwanger}}
  (\bibinfo{year}{2011}), \eprint{1108.0157}.

\bibitem[{\citenamefont{Alvarez-Gaume et~al.}(1983)\citenamefont{Alvarez-Gaume,
  Polchinski, and Wise}}]{AlvarezGaume:1983gj}
\bibinfo{author}{\bibfnamefont{L.}~\bibnamefont{Alvarez-Gaume}},
  \bibinfo{author}{\bibfnamefont{J.}~\bibnamefont{Polchinski}},
  \bibnamefont{and} \bibinfo{author}{\bibfnamefont{M.~B.} \bibnamefont{Wise}},
  \bibinfo{journal}{Nucl. Phys.} \textbf{\bibinfo{volume}{B221}},
  \bibinfo{pages}{495} (\bibinfo{year}{1983}).

\bibitem[{\citenamefont{Kounnas et~al.}(1984)\citenamefont{Kounnas, Lahanas,
  Nanopoulos, and Quiros}}]{Kounnas:1983td}
\bibinfo{author}{\bibfnamefont{C.}~\bibnamefont{Kounnas}},
  \bibinfo{author}{\bibfnamefont{A.~B.} \bibnamefont{Lahanas}},
  \bibinfo{author}{\bibfnamefont{D.~V.} \bibnamefont{Nanopoulos}},
  \bibnamefont{and} \bibinfo{author}{\bibfnamefont{M.}~\bibnamefont{Quiros}},
  \bibinfo{journal}{Nucl. Phys.} \textbf{\bibinfo{volume}{B236}},
  \bibinfo{pages}{438} (\bibinfo{year}{1984}).

\bibitem[{\citenamefont{Claudson et~al.}(1983)\citenamefont{Claudson, Hall, and
  Hinchliffe}}]{Claudson:1983et}
\bibinfo{author}{\bibfnamefont{M.}~\bibnamefont{Claudson}},
  \bibinfo{author}{\bibfnamefont{L.~J.} \bibnamefont{Hall}}, \bibnamefont{and}
  \bibinfo{author}{\bibfnamefont{I.}~\bibnamefont{Hinchliffe}},
  \bibinfo{journal}{Nucl. Phys.} \textbf{\bibinfo{volume}{B228}},
  \bibinfo{pages}{501} (\bibinfo{year}{1983}).

\bibitem[{\citenamefont{Drees et~al.}(1985)\citenamefont{Drees, Gluck, and
  Grassie}}]{Drees:1985ie}
\bibinfo{author}{\bibfnamefont{M.}~\bibnamefont{Drees}},
  \bibinfo{author}{\bibfnamefont{M.}~\bibnamefont{Gluck}}, \bibnamefont{and}
  \bibinfo{author}{\bibfnamefont{K.}~\bibnamefont{Grassie}},
  \bibinfo{journal}{Phys. Lett.} \textbf{\bibinfo{volume}{B157}},
  \bibinfo{pages}{164} (\bibinfo{year}{1985}).

\bibitem[{\citenamefont{Gunion et~al.}(1988)\citenamefont{Gunion, Haber, and
  Sher}}]{Gunion:1987qv}
\bibinfo{author}{\bibfnamefont{J.~F.} \bibnamefont{Gunion}},
  \bibinfo{author}{\bibfnamefont{H.~E.} \bibnamefont{Haber}}, \bibnamefont{and}
  \bibinfo{author}{\bibfnamefont{M.}~\bibnamefont{Sher}},
  \bibinfo{journal}{Nucl. Phys.} \textbf{\bibinfo{volume}{B306}},
  \bibinfo{pages}{1} (\bibinfo{year}{1988}).

\bibitem[{\citenamefont{Komatsu}(1988)}]{Komatsu:1988mt}
\bibinfo{author}{\bibfnamefont{H.}~\bibnamefont{Komatsu}},
  \bibinfo{journal}{Phys. Lett.} \textbf{\bibinfo{volume}{B215}},
  \bibinfo{pages}{323} (\bibinfo{year}{1988}).

\bibitem[{\citenamefont{Gamberini et~al.}(1990)\citenamefont{Gamberini,
  Ridolfi, and Zwirner}}]{Gamberini:1989jw}
\bibinfo{author}{\bibfnamefont{G.}~\bibnamefont{Gamberini}},
  \bibinfo{author}{\bibfnamefont{G.}~\bibnamefont{Ridolfi}}, \bibnamefont{and}
  \bibinfo{author}{\bibfnamefont{F.}~\bibnamefont{Zwirner}},
  \bibinfo{journal}{Nucl. Phys.} \textbf{\bibinfo{volume}{B331}},
  \bibinfo{pages}{331} (\bibinfo{year}{1990}).

\bibitem[{\citenamefont{Casas et~al.}(1996)\citenamefont{Casas, Lleyda, and
  Munoz}}]{Casas:1995pd}
\bibinfo{author}{\bibfnamefont{J.~A.} \bibnamefont{Casas}},
  \bibinfo{author}{\bibfnamefont{A.}~\bibnamefont{Lleyda}}, \bibnamefont{and}
  \bibinfo{author}{\bibfnamefont{C.}~\bibnamefont{Munoz}},
  \bibinfo{journal}{Nucl. Phys.} \textbf{\bibinfo{volume}{B471}},
  \bibinfo{pages}{3} (\bibinfo{year}{1996}), \eprint{hep-ph/9507294}.

\bibitem[{\citenamefont{Kobayashi and Shimomura}(2010)}]{Kobayashi:2010zx}
\bibinfo{author}{\bibfnamefont{T.}~\bibnamefont{Kobayashi}} \bibnamefont{and}
  \bibinfo{author}{\bibfnamefont{T.}~\bibnamefont{Shimomura}},
  \bibinfo{journal}{Phys. Rev.} \textbf{\bibinfo{volume}{D82}},
  \bibinfo{pages}{035008} (\bibinfo{year}{2010}), \eprint{1006.0062}.

\bibitem[{\citenamefont{Kanehata et~al.}(2010)\citenamefont{Kanehata,
  Kobayashi, Konishi, and Shimomura}}]{Kanehata:2010ci}
\bibinfo{author}{\bibfnamefont{Y.}~\bibnamefont{Kanehata}},
  \bibinfo{author}{\bibfnamefont{T.}~\bibnamefont{Kobayashi}},
  \bibinfo{author}{\bibfnamefont{Y.}~\bibnamefont{Konishi}}, \bibnamefont{and}
  \bibinfo{author}{\bibfnamefont{T.}~\bibnamefont{Shimomura}},
  \bibinfo{journal}{Phys. Rev.} \textbf{\bibinfo{volume}{D82}},
  \bibinfo{pages}{075018} (\bibinfo{year}{2010}), \eprint{1008.0593}.

\bibitem[{\citenamefont{Ellwanger et~al.}(1997)\citenamefont{Ellwanger,
  Rausch~de Traubenberg, and Savoy}}]{Ellwanger:1996gw}
\bibinfo{author}{\bibfnamefont{U.}~\bibnamefont{Ellwanger}},
  \bibinfo{author}{\bibfnamefont{M.}~\bibnamefont{Rausch~de Traubenberg}},
  \bibnamefont{and} \bibinfo{author}{\bibfnamefont{C.~A.} \bibnamefont{Savoy}},
  \bibinfo{journal}{Nucl. Phys.} \textbf{\bibinfo{volume}{B492}},
  \bibinfo{pages}{21} (\bibinfo{year}{1997}), \eprint{hep-ph/9611251}.

\bibitem[{\citenamefont{Funakubo and Tao}(2005)}]{Funakubo:2004ka}
\bibinfo{author}{\bibfnamefont{K.}~\bibnamefont{Funakubo}} \bibnamefont{and}
  \bibinfo{author}{\bibfnamefont{S.}~\bibnamefont{Tao}},
  \bibinfo{journal}{Prog. Theor. Phys.} \textbf{\bibinfo{volume}{113}},
  \bibinfo{pages}{821} (\bibinfo{year}{2005}), \eprint{hep-ph/0409294}.

\bibitem[{\citenamefont{Cerdeno et~al.}(2004)\citenamefont{Cerdeno, Hugonie,
  Lopez-Fogliani, Munoz, and Teixeira}}]{Cerdeno:2004xw}
\bibinfo{author}{\bibfnamefont{D.~G.} \bibnamefont{Cerdeno}},
  \bibinfo{author}{\bibfnamefont{C.}~\bibnamefont{Hugonie}},
  \bibinfo{author}{\bibfnamefont{D.~E.} \bibnamefont{Lopez-Fogliani}},
  \bibinfo{author}{\bibfnamefont{C.}~\bibnamefont{Munoz}}, \bibnamefont{and}
  \bibinfo{author}{\bibfnamefont{A.~M.} \bibnamefont{Teixeira}},
  \bibinfo{journal}{JHEP} \textbf{\bibinfo{volume}{12}}, \bibinfo{pages}{048}
  (\bibinfo{year}{2004}), \eprint{hep-ph/0408102}.

\bibitem[{\citenamefont{Maniatis et~al.}(2007)\citenamefont{Maniatis, von
  Manteuffel, and Nachtmann}}]{Maniatis:2006jd}
\bibinfo{author}{\bibfnamefont{M.}~\bibnamefont{Maniatis}},
  \bibinfo{author}{\bibfnamefont{A.}~\bibnamefont{von Manteuffel}},
  \bibnamefont{and}
  \bibinfo{author}{\bibfnamefont{O.}~\bibnamefont{Nachtmann}},
  \bibinfo{journal}{Eur. Phys. J.} \textbf{\bibinfo{volume}{C49}},
  \bibinfo{pages}{1067} (\bibinfo{year}{2007}), \eprint{hep-ph/0608314}.

\bibitem[{\citenamefont{Cerdeno et~al.}(2007)\citenamefont{Cerdeno, Gabrielli,
  Lopez-Fogliani, Munoz, and Teixeira}}]{Cerdeno:2007sn}
\bibinfo{author}{\bibfnamefont{D.~G.} \bibnamefont{Cerdeno}},
  \bibinfo{author}{\bibfnamefont{E.}~\bibnamefont{Gabrielli}},
  \bibinfo{author}{\bibfnamefont{D.~E.} \bibnamefont{Lopez-Fogliani}},
  \bibinfo{author}{\bibfnamefont{C.}~\bibnamefont{Munoz}}, \bibnamefont{and}
  \bibinfo{author}{\bibfnamefont{A.~M.} \bibnamefont{Teixeira}},
  \bibinfo{journal}{JCAP} \textbf{\bibinfo{volume}{0706}}, \bibinfo{pages}{008}
  (\bibinfo{year}{2007}), \eprint{hep-ph/0701271}.

\bibitem[{\citenamefont{Franceschini and Gori}(2011)}]{Franceschini:2010qz}
\bibinfo{author}{\bibfnamefont{R.}~\bibnamefont{Franceschini}}
  \bibnamefont{and} \bibinfo{author}{\bibfnamefont{S.}~\bibnamefont{Gori}},
  \bibinfo{journal}{JHEP} \textbf{\bibinfo{volume}{05}}, \bibinfo{pages}{084}
  (\bibinfo{year}{2011}), \eprint{1005.1070}.

\bibitem[{\citenamefont{Cerdeno et~al.}(2011)\citenamefont{Cerdeno, Huh, Peiro,
  and Seto}}]{Cerdeno:2011qv}
\bibinfo{author}{\bibfnamefont{D.~G.} \bibnamefont{Cerdeno}},
  \bibinfo{author}{\bibfnamefont{J.-H.} \bibnamefont{Huh}},
  \bibinfo{author}{\bibfnamefont{M.}~\bibnamefont{Peiro}}, \bibnamefont{and}
  \bibinfo{author}{\bibfnamefont{O.}~\bibnamefont{Seto}},
  \bibinfo{journal}{JCAP} \textbf{\bibinfo{volume}{1111}}, \bibinfo{pages}{027}
  (\bibinfo{year}{2011}), \eprint{1108.0978}.

\bibitem[{\citenamefont{Kanehata et~al.}(2011)\citenamefont{Kanehata,
  Kobayashi, Konishi, Seto, and Shimomura}}]{Kanehata:2011ei}
\bibinfo{author}{\bibfnamefont{Y.}~\bibnamefont{Kanehata}},
  \bibinfo{author}{\bibfnamefont{T.}~\bibnamefont{Kobayashi}},
  \bibinfo{author}{\bibfnamefont{Y.}~\bibnamefont{Konishi}},
  \bibinfo{author}{\bibfnamefont{O.}~\bibnamefont{Seto}}, \bibnamefont{and}
  \bibinfo{author}{\bibfnamefont{T.}~\bibnamefont{Shimomura}},
  \bibinfo{journal}{Prog. Theor. Phys.} \textbf{\bibinfo{volume}{126}},
  \bibinfo{pages}{1051} (\bibinfo{year}{2011}), \eprint{1103.5109}.

\bibitem[{\citenamefont{Bertuzzo and Farina}(2012)}]{Bertuzzo:2011ij}
\bibinfo{author}{\bibfnamefont{E.}~\bibnamefont{Bertuzzo}} \bibnamefont{and}
  \bibinfo{author}{\bibfnamefont{M.}~\bibnamefont{Farina}},
  \bibinfo{journal}{Phys. Rev.} \textbf{\bibinfo{volume}{D85}},
  \bibinfo{pages}{015011} (\bibinfo{year}{2012}), \eprint{1105.5389}.

\bibitem[{\citenamefont{Hamaguchi et~al.}(2011)\citenamefont{Hamaguchi,
  Nakayama, and Yokozaki}}]{Hamaguchi:2011kt}
\bibinfo{author}{\bibfnamefont{K.}~\bibnamefont{Hamaguchi}},
  \bibinfo{author}{\bibfnamefont{K.}~\bibnamefont{Nakayama}}, \bibnamefont{and}
  \bibinfo{author}{\bibfnamefont{N.}~\bibnamefont{Yokozaki}}
  (\bibinfo{year}{2011}), \eprint{1111.1601}.

\bibitem[{\citenamefont{Ellwanger and Hugonie}(2006)}]{Ellwanger:2005dv}
\bibinfo{author}{\bibfnamefont{U.}~\bibnamefont{Ellwanger}} \bibnamefont{and}
  \bibinfo{author}{\bibfnamefont{C.}~\bibnamefont{Hugonie}},
  \bibinfo{journal}{Comput. Phys. Commun.} \textbf{\bibinfo{volume}{175}},
  \bibinfo{pages}{290} (\bibinfo{year}{2006}), \eprint{hep-ph/0508022}.

\bibitem[{\citenamefont{Ellwanger and
  Hugonie}(2007{\natexlab{a}})}]{Ellwanger:2006rn}
\bibinfo{author}{\bibfnamefont{U.}~\bibnamefont{Ellwanger}} \bibnamefont{and}
  \bibinfo{author}{\bibfnamefont{C.}~\bibnamefont{Hugonie}},
  \bibinfo{journal}{Comput. Phys. Commun.} \textbf{\bibinfo{volume}{177}},
  \bibinfo{pages}{399} (\bibinfo{year}{2007}{\natexlab{a}}),
  \eprint{hep-ph/0612134}.

\bibitem[{\citenamefont{Djouadi et~al.}(2008)}]{Djouadi:2008uw}
\bibinfo{author}{\bibfnamefont{A.}~\bibnamefont{Djouadi}} \bibnamefont{et~al.},
  \bibinfo{journal}{JHEP} \textbf{\bibinfo{volume}{07}}, \bibinfo{pages}{002}
  (\bibinfo{year}{2008}), \eprint{0801.4321}.

\bibitem[{\citenamefont{Romao}(1986)}]{Romao:1986jy}
\bibinfo{author}{\bibfnamefont{J.~C.} \bibnamefont{Romao}},
  \bibinfo{journal}{Phys. Lett.} \textbf{\bibinfo{volume}{B173}},
  \bibinfo{pages}{309} (\bibinfo{year}{1986}).

\bibitem[{\citenamefont{Ellwanger and
  Hugonie}(2007{\natexlab{b}})}]{Ellwanger:2006rm}
\bibinfo{author}{\bibfnamefont{U.}~\bibnamefont{Ellwanger}} \bibnamefont{and}
  \bibinfo{author}{\bibfnamefont{C.}~\bibnamefont{Hugonie}},
  \bibinfo{journal}{Mod. Phys. Lett.} \textbf{\bibinfo{volume}{A22}},
  \bibinfo{pages}{1581} (\bibinfo{year}{2007}{\natexlab{b}}),
  \eprint{hep-ph/0612133}.

\bibitem[{\citenamefont{Falck}(1986)}]{Falck:1985aa}
\bibinfo{author}{\bibfnamefont{N.~K.} \bibnamefont{Falck}},
  \bibinfo{journal}{Z. Phys.} \textbf{\bibinfo{volume}{C30}},
  \bibinfo{pages}{247} (\bibinfo{year}{1986}).

\bibitem[{\citenamefont{Miller et~al.}(2004)\citenamefont{Miller, Nevzorov, and
  Zerwas}}]{Miller:2003ay}
\bibinfo{author}{\bibfnamefont{D.~J.} \bibnamefont{Miller}, \bibfnamefont{2}},
  \bibinfo{author}{\bibfnamefont{R.}~\bibnamefont{Nevzorov}}, \bibnamefont{and}
  \bibinfo{author}{\bibfnamefont{P.~M.} \bibnamefont{Zerwas}},
  \bibinfo{journal}{Nucl. Phys.} \textbf{\bibinfo{volume}{B681}},
  \bibinfo{pages}{3} (\bibinfo{year}{2004}), \eprint{hep-ph/0304049}.

\bibitem[{\citenamefont{Barbieri et~al.}(2007)\citenamefont{Barbieri, Hall,
  Nomura, and Rychkov}}]{Barbieri:2006bg}
\bibinfo{author}{\bibfnamefont{R.}~\bibnamefont{Barbieri}},
  \bibinfo{author}{\bibfnamefont{L.~J.} \bibnamefont{Hall}},
  \bibinfo{author}{\bibfnamefont{Y.}~\bibnamefont{Nomura}}, \bibnamefont{and}
  \bibinfo{author}{\bibfnamefont{V.~S.} \bibnamefont{Rychkov}},
  \bibinfo{journal}{Phys. Rev.} \textbf{\bibinfo{volume}{D75}},
  \bibinfo{pages}{035007} (\bibinfo{year}{2007}), \eprint{hep-ph/0607332}.

\bibitem[{\citenamefont{Lodone}(2011)}]{Lodone:2011ax}
\bibinfo{author}{\bibfnamefont{P.}~\bibnamefont{Lodone}},
  \bibinfo{journal}{Int.J.Mod.Phys.} \textbf{\bibinfo{volume}{A26}},
  \bibinfo{pages}{4053} (\bibinfo{year}{2011}), \eprint{1105.5248}.

\bibitem[{\citenamefont{Bertuzzo and Farina}(2011)}]{Bertuzzo:2011aa}
\bibinfo{author}{\bibfnamefont{E.}~\bibnamefont{Bertuzzo}} \bibnamefont{and}
  \bibinfo{author}{\bibfnamefont{M.}~\bibnamefont{Farina}}
  (\bibinfo{year}{2011}), \eprint{1112.2190}.

\bibitem[{\citenamefont{Coleman}(1977{\natexlab{a}})}]{Coleman:1977py}
\bibinfo{author}{\bibfnamefont{S.~R.} \bibnamefont{Coleman}},
  \bibinfo{journal}{Phys. Rev.} \textbf{\bibinfo{volume}{D15}},
  \bibinfo{pages}{2929} (\bibinfo{year}{1977}{\natexlab{a}}).

\bibitem[{\citenamefont{Coleman}(1977{\natexlab{b}})}]{Coleman:1977py2}
\bibinfo{author}{\bibfnamefont{S.~R.} \bibnamefont{Coleman}},
  \bibinfo{journal}{[Erratum-ibid] Phys. Rev.} \textbf{\bibinfo{volume}{D16}},
  \bibinfo{pages}{1248} (\bibinfo{year}{1977}{\natexlab{b}}).

\bibitem[{\citenamefont{Callan and Coleman}(1977)}]{Callan:1977pt}
\bibinfo{author}{\bibfnamefont{C.~G.} \bibnamefont{Callan}, \bibfnamefont{Jr.}}
  \bibnamefont{and} \bibinfo{author}{\bibfnamefont{S.~R.}
  \bibnamefont{Coleman}}, \bibinfo{journal}{Phys. Rev.}
  \textbf{\bibinfo{volume}{D16}}, \bibinfo{pages}{1762} (\bibinfo{year}{1977}).

\bibitem[{\citenamefont{Coleman and De~Luccia}(1980)}]{Coleman:1980aw}
\bibinfo{author}{\bibfnamefont{S.~R.} \bibnamefont{Coleman}} \bibnamefont{and}
  \bibinfo{author}{\bibfnamefont{F.}~\bibnamefont{De~Luccia}},
  \bibinfo{journal}{Phys. Rev.} \textbf{\bibinfo{volume}{D21}},
  \bibinfo{pages}{3305} (\bibinfo{year}{1980}).

\bibitem[{\citenamefont{Lee and Weinberg}(1986)}]{Lee:1985uv}
\bibinfo{author}{\bibfnamefont{K.-M.} \bibnamefont{Lee}} \bibnamefont{and}
  \bibinfo{author}{\bibfnamefont{E.~J.} \bibnamefont{Weinberg}},
  \bibinfo{journal}{Nucl. Phys.} \textbf{\bibinfo{volume}{B267}},
  \bibinfo{pages}{181} (\bibinfo{year}{1986}).

\bibitem[{\citenamefont{Duncan and Jensen}(1992)}]{Duncan:1992ai}
\bibinfo{author}{\bibfnamefont{M.~J.} \bibnamefont{Duncan}} \bibnamefont{and}
  \bibinfo{author}{\bibfnamefont{L.~G.} \bibnamefont{Jensen}},
  \bibinfo{journal}{Phys. Lett.} \textbf{\bibinfo{volume}{B291}},
  \bibinfo{pages}{109} (\bibinfo{year}{1992}).

\end{thebibliography}

\end{document}